\documentclass[ALICE,manyauthors]{cernphprep}

\usepackage[comma,square,numbers,sort&compress]{natbib}
\usepackage{hyperref}
\usepackage{lineno}

\usepackage[T1]{fontenc} 

\usepackage{pifont}
\usepackage{enumitem}
\usepackage{longtable}

\usepackage{subfigure}

\usepackage{wrapfig}
\usepackage{dcolumn}
\usepackage{longtable}
\usepackage[percent]{overpic}

\usepackage[normalem]{ulem} 

\RequirePackage{lineno}
\makeatletter
\newcommand{\thickhline}{%
    \noalign {\ifnum 0=`}\fi \hrule height 1pt
    \futurelet \reserved@a \@xhline
}
\newcolumntype{"}{@{\hskip\tabcolsep\vrule width 1pt\hskip\tabcolsep}}
\makeatother

\def\bc{\begin{center}}
\def\ec{\end{center}}
\def\beq{\begin{equation}}
\def\eeq{\end{equation}}

\def\av#1{\langle #1 \rangle}

\def\$p_t${$p_{\rm t}$}

\def\bcor{b_{\rm corr}}

\def\nB{n^{}_{\rm B}}
\def\nF{n^{}_{\rm F}}

\def\avr#1#2{\langle {#1} \rangle^{}_{#2}}

\def\nF{{n^{}_{\rm F}}}
\def\nB{{n^{}_{\rm B}}}

\def\bc{\begin{center}}
\def\ec{\end{center}}
\def\beq{\begin{equation}}
\def\eeq{\end{equation}}

\def\av#1{\langle #1 \rangle}
\def\avr#1#2{\langle {#1} \rangle^{}_{#2}}

\def\eg{{\eta^{}_{\rm gap}}}

\def\pt{p_{\rm{T}}}
\def\ptmin{p_{\rm{T min}}}
\def\ptmin{p_{\rm T}^{\rm min}}
\def\ptmax{p_{\rm T}^{\rm max}}

\def\nF{{n_{\rm F}^{}}}

\def\nB{{n_{\rm B}^{}}}

\def\deta{{\delta \eta}}

\def\dphi{{\delta \varphi}}

\def\bc{\begin{center}}
\def\ec{\end{center}}
\def\beq{\begin{equation}}
\def\eeq{\end{equation}}

\def\av#1{\langle #1 \rangle}

\def\avr#1#2{\langle {#1} \rangle^{}_{#2}}
\def\nF{{n_{\rm F}^{}}}

\def\nB{{n_{\rm B}^{}}}

\def\Dy{{\delta\eta}}

\def\eg{{\eta^{}_{\rm gap}}}

\def\esep{{\eta^{}_{\rm sep}}}
\def\fsep{{\varphi^{}_{\rm sep}}}

\def\sigFB{{\sigma^2_{\nF+\nB}}}
\def\sigF{{\sigma^2_{\nF}}}
\def\sigB{{\sigma^2_{\nB}}}
\def\sigN{{\sigma^2_N}}

\begin{document}
\begin{titlepage}
%

\PHyear{2015}
\PHnumber{012} 
\PHdate{21 January}  
%
%
\title{Forward-backward multiplicity correlations \\
in pp collisions at $\mathbf{\sqrt{s}}$\ =\ 0.9, 2.76 and 7 TeV}
\ShortTitle{Forward-backward multiplicity correlations 
in pp collisions  
}   
%
\Collaboration{ALICE Collaboration%
\thanks{See Appendix~\ref{app:collab} for the list of collaboration
members}}
\ShortAuthor{ALICE Collaboration}      
%


\begin{abstract}

The strength of forward-backward (FB) multiplicity correlations is measured by the~ALICE detector in proton-proton (pp) collisions 
at $\sqrt{s}$\ =\ 0.9, 2.76 and 7~TeV.
The measurement is performed in the central pseudorapidity region ($|\eta| < 0.8$) for the transverse momentum $\pt>0.3$~GeV/$c$.
Two separate pseudorapidity windows of width ($\delta \eta$) ranging from 0.2  to 0.8 are chosen  symmetrically around $\eta$~=~0.
The multiplicity correlation strength ($\bcor$) is studied as a function of the pseudorapidity gap ($\eg$) between the two windows 
as well as the width of these windows.
The correlation strength is found to decrease with increasing $\eg$ and shows a non-linear increase with $\delta \eta$.
A sizable increase of the correlation strength with the collision energy, which cannot be explained  exclusively by the increase 
of the mean multiplicity inside the windows, is observed.
The correlation coefficient is also measured for multiplicities 
in different configurations of two azimuthal sectors selected within the symmetric FB $\eta$-windows.
Two different contributions, the short-range (SR) and the long-range (LR),  are observed.
The energy dependence of $\bcor$ is found to be weak for the SR component while  it is strong for the LR component.
Moreover, the correlation coefficient is studied for particles belonging to various transverse momentum intervals chosen to have 
the same mean multiplicity.  
Both  SR and LR contributions to $\bcor$ are found  to increase with $\pt$ in this case.
Results are compared to PYTHIA and PHOJET event generators and to a string-based phenomenological model. The observed dependencies of $\bcor$
add new constraints on  phenomenological models.

\end{abstract}
\end{titlepage}
\setcounter{page}{2}

\section{Introduction}
\label{sec:introduction}

We report a detailed study of correlations between multiplicities in pp collisions at 0.9,  2.76 and 7~TeV.
The correlations are obtained from event-by-event multiplicity measurements in pseudorapidity
 ($\eta$) and azimuth ($\varphi$) separated intervals.
The  intervals are selected one in the forward and another in the backward hemispheres
in the center-of-mass system, therefore the correlations are referred to  as forward-backward (FB) correlations.

The FB correlation strength is characterized by the correlation coefficient,
$\bcor$, which is obtained from a linear regression analysis of the average multiplicity
measured in the backward rapidity hemisphere ($\avr\nB\nF$) 
as a function of the event multiplicity in the forward hemisphere ($\nF$):
\begin{equation}\label{linear}
\avr\nB\nF = a + \bcor\cdot\nF \  .
\end{equation}
This linear relation
\eqref{linear}
has been observed experimentally
\cite{bLinear, capella1994,ua5,Uhlig-78}
and is discussed in \cite{LRC, Fowler-88, CapellaKrzywicki}.
Under the assumption of linear correlation between $\nF$ and $\nB$, the 
Pearson correlation coefficient
\begin{equation}
\label{b-alt}
\bcor = \frac{\av{\nB\nF}-\av{\nB}\av{\nF}} {\av{n_{\rm F}^2}-\av{\nF}^2_{}}  
\end{equation}
can be used for the experimental determination of $\bcor$ \cite{capella1994}.
Since the parameter $a$ is given by
$a = \av{\nB} - \bcor  \av{\nF} $, 
it adds no additional  information and usually is not considered  \cite{LRC, Fowler-88, CapellaKrzywicki}.

Heretofore, FB multiplicity correlations were studied experimentally in a large number of collision systems 
including
$e^{+}e^{-}$, $\mu^{+}{\rm p}$, pp, ${\rm p}\overline{\rm p}$ and A--A interactions
\cite{LRCee, LRCee2, LRCee3, LRCpp, Uhlig-78, 
LRCpp4, ISR, ua5}.
No FB multiplicity correlations were observed in $e^{+}e^{-}$ annihilation at  $\sqrt{s}=29$ GeV.
This was interpreted as the consequence of independent fragmentation of the forward and backward jets produced in this process
\cite{Derrick}.
In contrast, in pp collisions at the ISR \cite{ISR}
at $\sqrt{s}=52.6$~GeV \cite{Uhlig-78} and in ${\rm p}\overline{\rm p}$  interactions at the ${\rm S p \overline{\rm p} S}$ collider
\cite{Alpgard} 
sizeable positive FB multiplicity correlations have been observed.
Their strength was found to increase strongly with  collision energy \cite{ua5},
which was confirmed later
at much higher energies ($\sqrt{s}\gtrsim 1$ TeV)
in ${\rm p}\overline{\rm p}$ collisions by the E735 collaboration at the Tevatron \cite{LRCpp4}
and  in pp collisions by the ATLAS experiment at the LHC ($\sqrt{s}=0.9$ and 7~TeV)
\cite{atlas-lrc}.
One of the observations reported by ATLAS is the decrease of  $\bcor$ 
with the increase of the minimum transverse momentum of charged particles.

The STAR collaboration at RHIC analysed the FB multiplicity correlations
in pp and Au--Au collisions at $\sqrt{s_{NN}} = 200$~GeV \cite{STAR-FBC}.
Strong correlation was observed in case of Au--Au collisions,
while in pp collisions $\bcor$ was found to be rather small ($\sim$0.1).
In the present paper we relate this to the use of
smaller pseudorapidity windows 
as compared to  previous pp and ${\rm p}\overline{\rm p}$ measurements.

Forward-backward multiplicity correlations in high energy pp and A--A collisions
also raise a considerable theoretical interest.
First attempts to explain this phenomenon 
\cite{CapellaKrzywicki,AK,BPV00,BDM07} were made
in the framework of  the Dual Parton Model (DPM) \cite{capella1994} and
the Quark Gluon String Model (QGSM) \cite{Kaid1, Kaid2}.
They provide  a quantitative description of multiparticle production in soft processes.
In improved versions of the models,  collectivity effects arising due to the interactions between strings,
which are particularly important in the case of A--A interactions,
were taken into account \cite{SRC-LRC-SFM,BP00,BKPV04,VK07}. 
These effects are based on the String Fusion Model (SFM) proposed in \cite{BP1,B-P}.
It was shown that 
these string interactions  
lead to a considerable modification of the FB correlation strength, along with the reduction of multiplicities, the increase of mean particle $\pt$,
and the enhancement of  heavy flavour production
in central A--A collisions \cite{SRC-LRC-SFM, STRANGE, MC-SFM}.

FB correlations are usually divided
into short and  long-range components \cite{CapellaKrzywicki,capella1994}.
In phenomenological models, 
short-range correlations (SRC) are assumed to be localized over a small range 
of  $\eta$-differences, up to one unit. They are induced by various short-range effects 
from single source fragmentation,
including particles produced from decays of clusters or resonances, jet and mini-jet induced correlations.
Long-range 
correlations (LRC) extend  over a wider range
in $\eta$. 
They originate from fluctuations in the number and properties of particle emitting sources (clusters, cut pomerons, strings, mini-jets etc.)
\cite{CapellaKrzywicki, capella1994, BPV00,
SRC-LRC-SFM, BP00, BKPV04, VK07}.

The SFM predicts that the variance of the number
of particle-emitting sources (strings) should be  damped by their fusion,
implying a reduction of multiplicity long-range correlations \cite{SRC-LRC-SFM,BKPV04,VK07}.
Contrary to this prediction, long-range correlations arising in the Color Glass Condensate model (CGC)  \cite{CGC} have been shown to increase with the centrality of the collision \cite{nestor}. 
Therefore, the investigation of correlations between various observables, measured in two different,
sufficiently separated  $\eta$-intervals, is considered  to be a powerful tool
for the exploration 
of the initial conditions of hadronic interactions \cite{Dumitru}. 
In the case of A--A collisions, these correlations induced across a wide range in $\eta$
are expected to reflect the earliest stages of the collisions, almost free from final state effects \cite{LRCCGC,nestor}.
The reference for the analysis of A--A collision dynamics can be obtained in pp collisions
by studying the dependence  of  FB correlations  on collision energy, particle pseudorapidity, azimuth and transverse momenta.

This paper is organized as follows:
Section \ref{sec:experimentalSetup}
provides experimental details, including 
the description of the procedures used for the event and track 
selection, the efficiency corrections
and systematic uncertainties estimates.
Sections \ref{sec:correlationsInEtaWithFullPhi} and \ref{sec:azimuthalExperimentalResults}
discuss the results on  FB  
multiplicity correlation measurements in $\eta$ in pp collisions at  $\sqrt{s}=0.9$, 2.76 and 7~TeV
and in $\eta$-$\phi$ windows
at   $\sqrt{s}=0.9$ and 7~TeV.
In Section \ref{sec:correlationsInEtaWithFullPhi}, we present 
dependences of the correlation coefficient on the gap between windows, their widths and the collision energy.
In Section \ref{sec:azimuthalExperimentalResults}, multiplicity correlations in windows separated in pseudorapidity and azimuth are studied,
and the comparison with Monte Carlo generators PYTHIA6 and PHOJET is discussed.
Results on multiplicity correlations in different $\pt$ ranges in pp collisions  at  $\sqrt{s}=7$~TeV are presented
in Section~\ref{sec:ptBinsStudy}.

\section{Data Analysis}
\label{sec:experimentalSetup}

\subsection{Experimental setup, event  and track selection}

The data presented in this paper were recorded with the ALICE detector~\cite{ALICE}
in pp collisions  at  $\sqrt{s}=0.9$, 2.76 and 7~TeV.
Charged primary particles are reconstructed 
with the central barrel detectors combining information 
from the Inner Tracking System (ITS) and the Time Projection Chamber (TPC).
Both detectors are located inside the  0.5~T solenoidal field. 

The ITS is composed of 3 different types of coordinate-sensitive  Si-detectors. It consists
of 2 silicon pixel innermost layers (SPD), 2 silicon drift (SDD) and 2 silicon strip (SSD) outer detector layers. 
The design allows for two-particle separation 
in events with multiplicity up to 
100 charged particles per ${\rm cm}^2$.
The SPD detector covers the pseudorapidity ranges $|\eta|<2$ for inner 
and $|\eta|<1.4$ for outer layers,
acceptances of SDD and SSD are $|\eta|<0.9$ and $|\eta|<1$, respectively. 
All ITS elements have a radiation length 
of about 1.1$\%$ $X_0$ per layer. 
The ITS provides reliable charged particle tracking 
down to transverse momenta of 0.1 GeV/$c$, ideal for the study of low-$\pt$ (soft) phenomena.

The ALICE TPC is the main tracking detector of the central rapidity region.
The TPC, together with the ITS, provides charged particle momentum
measurement, particle identification and vertex
determination with good 
momentum and ${\rm d}E/{\rm d}x$  resolution as well as two-track separation 
of identified hadrons and leptons  in the $\pt$ region below  10 GeV/$c$.
The TPC has an acceptance of $|\eta| < 0.9$
for tracks which reach the outer radius of the TPC and up to
$|\eta| < 1.5$ 
for tracks that exit through the endcap of the TPC.

For the present analysis, minimum bias pp events are used.
The minimum-bias trigger required a hit in one of the
forward scintillator counters (VZERO) or in one of the two SPD layers.
The VZERO timing signal
was used to reject beam-gas and beam-halo collisions.
The primary vertex was reconstructed using
the combined track information from the TPC and ITS, 
and only events with primary vertices
lying within $\pm 10$~cm from the centre of the apparatus 
are selected. 
In this way a uniform  acceptance in the central
pseudorapidity region $|\eta| < 0.8$ is ensured.
The data samples   for  $\sqrt{s}=0.9$, 2.76 and 7~TeV comprise  $2\times10^6$,   $10\times10^6$, and $6.5\times10^6$  events, respectively.
Only runs with low probability to produce several separate events per one  bunch crossing (so-called pile-up events)
were used in this analysis.

To obtain high  
tracking efficiency and to reduce efficiency losses due to detector
boundaries, tracks are selected with 
$\pt>$ 0.3\,GeV/{\it c} in the pseudorapidity range $|\eta| < 0.8$. 
Employing a Kalman filter technique, tracks are reconstructed using
space-time points measured by the TPC.
Tracks with at least 70 space-points associated and track fitting  $\chi^2/n_{\rm dof}$ less than 2 are accepted.
Additionally, at least two hits in the ITS must be associated
with the track. Tracks are also rejected 
if their distance of closest approach (DCA) to the reconstructed event vertex 
is larger than 0.3~cm in either the transverse or the longitudinal plane.
For the chosen selection criteria,
the tracking efficiency for charged particles with $\pt>0.3$~GeV/$c$
is  about 80\%.

\begin{figure}[t!f]
\centering
\begin{minipage}{0.48\textwidth}
\centering
\begin{overpic}[width=0.96\textwidth]{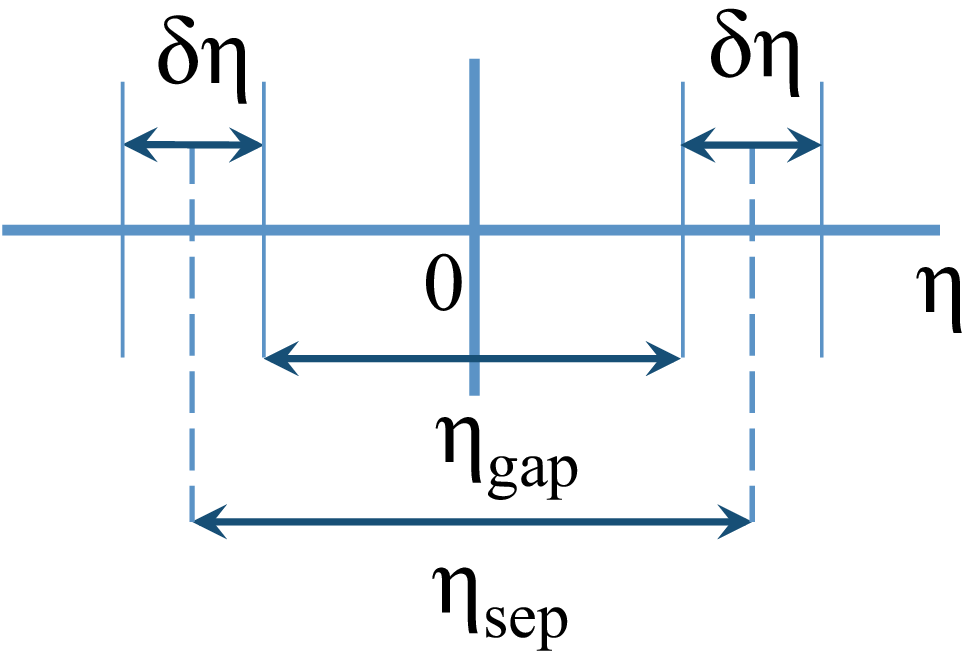}
\end{overpic}
\caption[Schematic representation of introduced variables $\deta$, $\eg$ and $\esep$.]
{Illustration of the variables $\deta$, $\eg$ and $\esep$ used in the present analysis.}
\label{fig:etaWindowsVarDefinitions}
\end{minipage}%
\quad
\begin{minipage}{0.48\textwidth}
\centering
\includegraphics[width=0.99\textwidth]{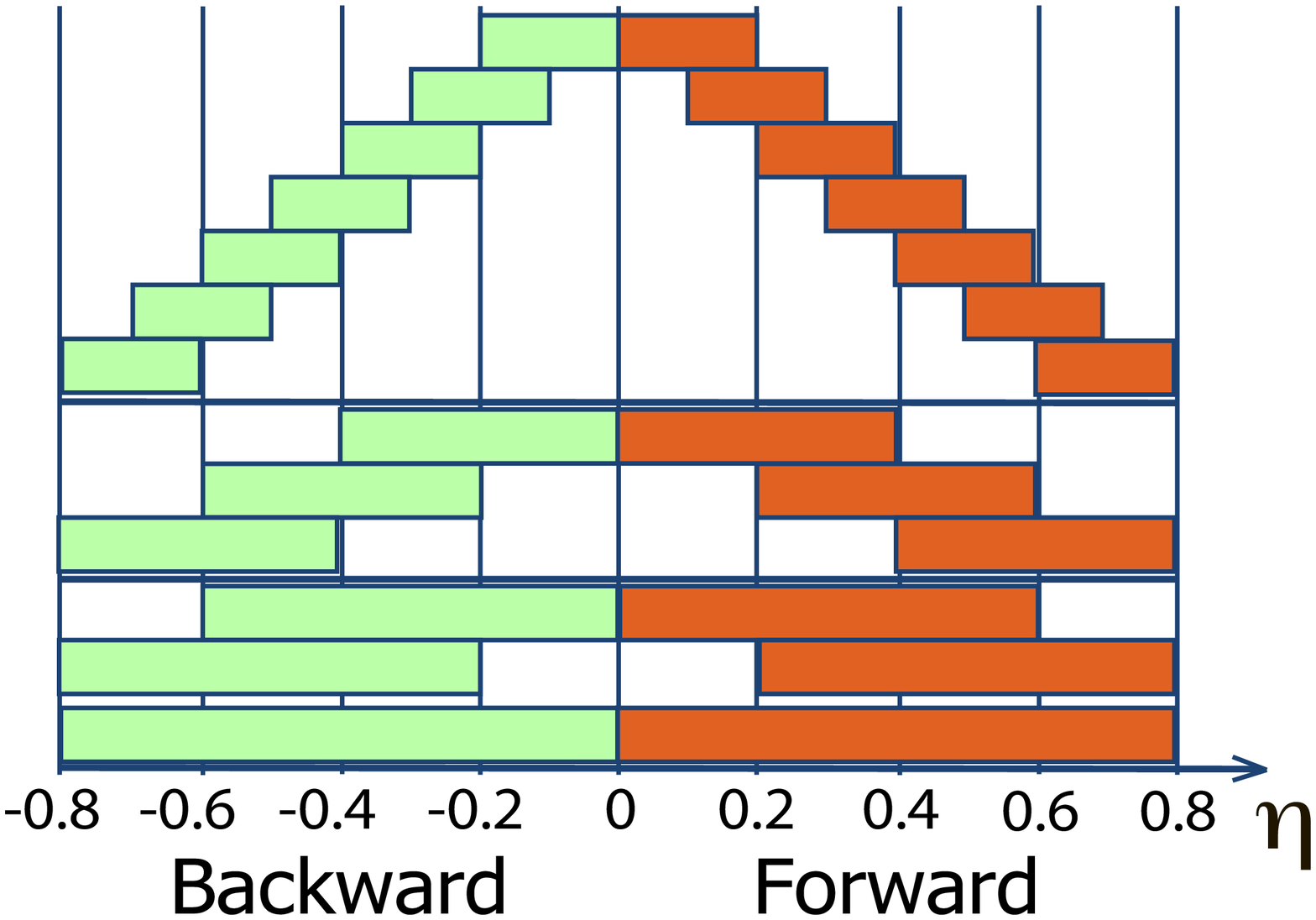}
\caption[Set of $\eta$-windows with chosen $\deta$  and $\eg$.]{
Illustration of sets  of $\eta$-windows with different widths $\deta$  and separation gaps $\eg$.}
\label{fig:etaWindowsView}
\end{minipage}
\end{figure}

\begin{figure}
\centering
\includegraphics[scale=1.2, angle=0]{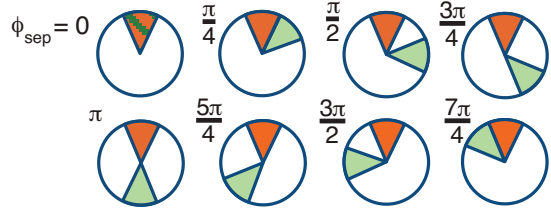} 
\caption{ Illustration of 8 configurations of  azimuthal sectors.  Forward and backward  pseudorapidity windows  
of the width $\deta=0.2$ 
are additionally split into 8 azimuthal  sectors with the width $\dphi$=$\pi$/4. 
The red sectors correspond to the first window of the FB pair, the green sectors to the second one.
The variable $\fsep$ is the separation  in azimuthal angle between centres of the sectors.
}
\label{fig:windowsPhiSectors}
\end{figure}

\subsection{Definition of counting windows}
\label{sec:etaPhiWindows}

Two intervals separated symmetrically around $\eta = 0$ with variable width $\deta$ ranging from 
0.2 to 0.8 are defined as ``forward'' (F, $\eta>0$) 
and ``backward'' (B, $\eta<0$) . 
Correlations between   multiplicities of charged particles ($n$) 
are studied as a function of the gap 
between the windows (denoted as \textbf{$\eg$}).
Another convenient variable is  $\esep$ which is the separation in pseudorapidity between centres of the windows.
These variables are illustrated in Fig.~\ref{fig:etaWindowsVarDefinitions}, 
and all configurations of window pairs chosen for the analysis are drawn in Fig.~\ref{fig:etaWindowsView}.

The analysis is extended to correlations between separated regions in the $\eta$-$\varphi$ plane (sectors).
The $\varphi$-angle space is split  into 8 sectors with the width $\dphi$=$\pi$/4 as shown in Fig.~\ref{fig:windowsPhiSectors}.
This selection is motivated by a compromise between granularity 
and statistical uncertainty. 
The definitions and equations, described  in Section \ref{sec:introduction},  
remain the same for the $\eta$-$\varphi$ windows. 
The acceptance of  the  windows is determined by their widths $\deta$ and $\dphi$
as the ALICE acceptance is approximately uniform in the selected ranges of $\eta$ and $\varphi$.

\subsection{Experimental procedures of the FB correlation coefficient measurement}
\label{sec:correlationMethods}

The present paper focuses on the study of FB correlation phenomena related to soft particle 
production. Therefore we restrict $\pt$ 
in $0.3<\pt<1.5$~GeV/$c$, 
except for the study of the $\pt$  dependence 
presented in Section~\ref{sec:ptBinsStudy}, where the $\pt$ range is $0.3<\pt<6$~GeV/$c$.

\begin{figure}[!t]
\centering
$\begin{array}{ccc}
\begin{overpic}[width=0.46\textwidth]{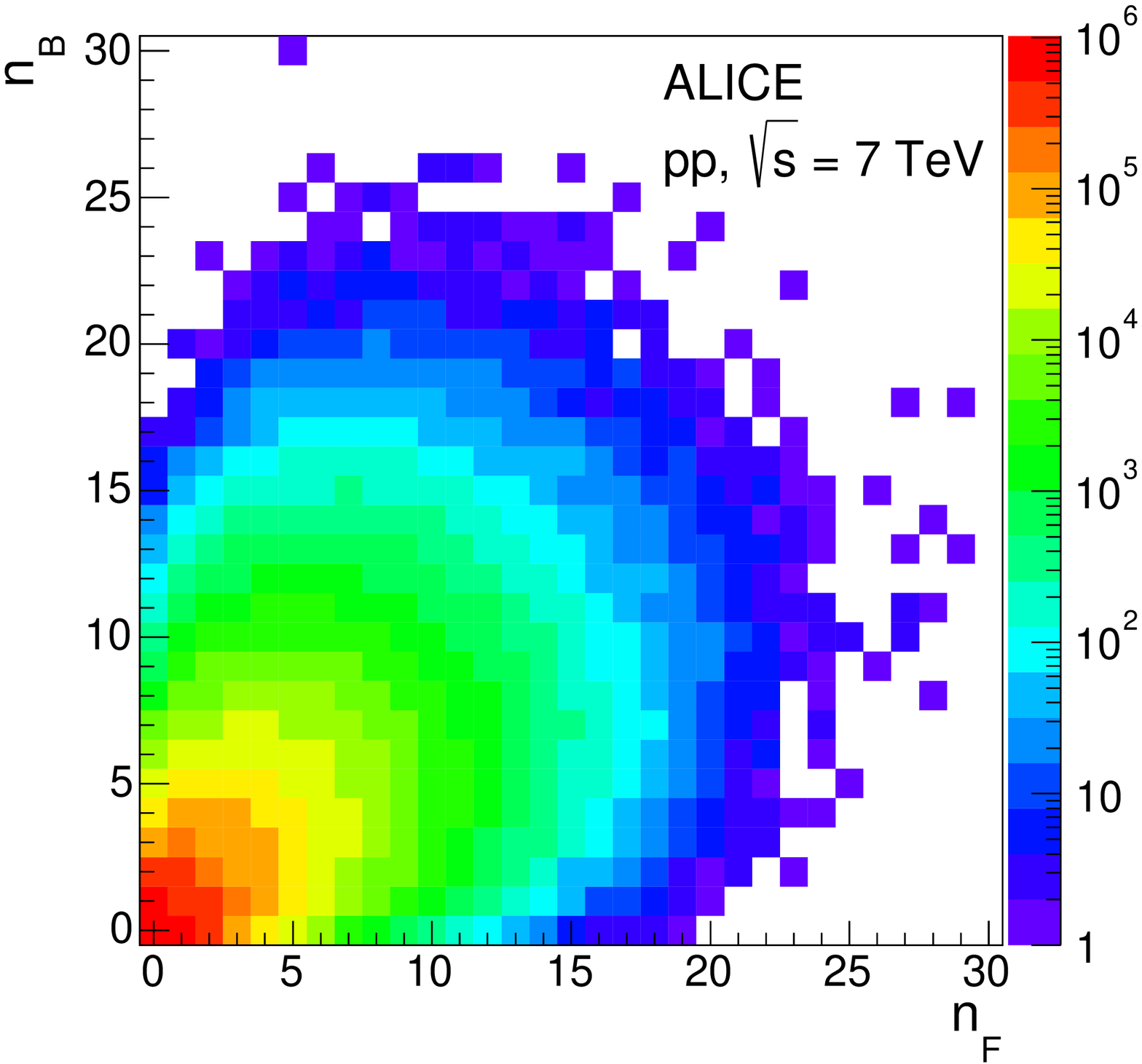}
\end{overpic}
&
\begin{overpic}[width=0.46\textwidth]{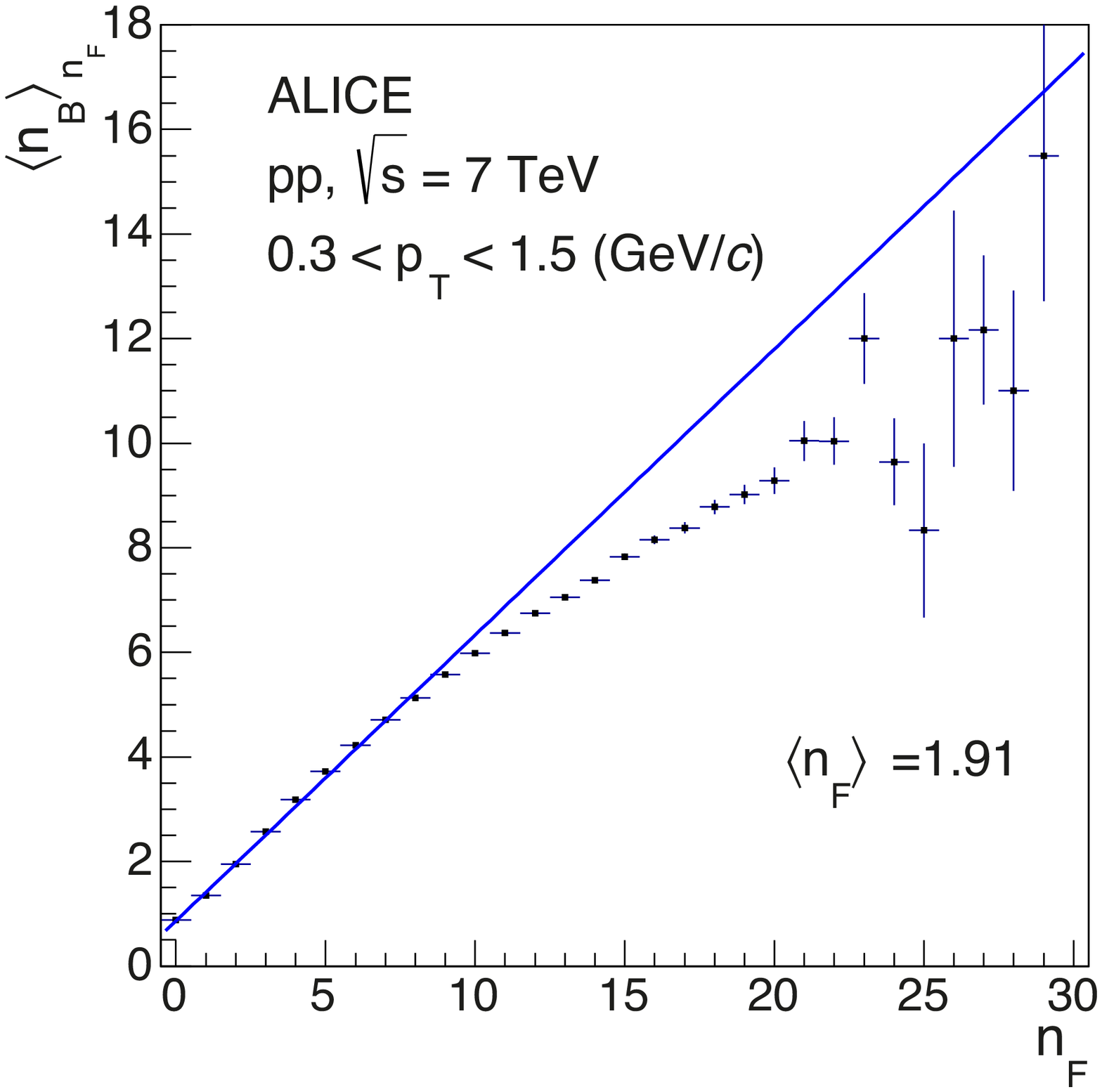}    
\end{overpic}
\end{array}$
\caption{
Illustrative example of 
forward versus backward raw multiplicity distribution for windows with $\deta=0.6$ and $\eg=0.4$ at $\sqrt{s}=7$~TeV (left) and corresponding correlation function (right). 
The correlation strength $\bcor$ is obtained from a linear fit 
according to Eq.~\ref{linear}. Since most of the statistics is at low multiplicities, 
the fit is mainly determined by the first points.
}
\label{fig:hist2D_and_profile}
\end{figure}

The correlation coefficients, $\bcor$, for each window pair can be calculated using two methods \cite{bLinear, capella1994,ua5,Uhlig-78}.
In the first method
values of $\av{\nB\nF}$, $\av{\nB}$, $\av{\nF}$ and $\av{n_F^2}$ 
are accumulated event-by-event and then $\bcor$ is determined 
using Eq.\ \ref{b-alt}.
In the second method,
$\bcor$ is calculated using linear regression. 
The 2-dimensional  distributions ($n_{\rm B}$,~$n_{\rm F}$) are
obtained integrating over all selected events, then the average backward multiplicity is calculated for each 
fixed value of the forward multiplicity,
and $\bcor$ is obtained from a linear fit to the correlation function
(see illustration in Fig.~\ref{fig:hist2D_and_profile}). 
Deviations from linear behavior 
may provide additional information, however, a detailed study of non-linearity in
the correlation function is beyond the scope of this paper.

It has been  shown 
that the results obtained with the two methods agree within statistical uncertainty.
In this work, results using the first method are presented.

\subsection{Corrections and systematic uncertainties }
\label{sec:correctionProcedures}

Acceptance and tracking efficiency corrections are extracted from Monte Carlo simulations using  
 PYTHIA6 \cite{pythia}  (Perugia 0 tune) and PHOJET \cite{phojet,phojet2} as particle generators 
followed by a full detector response simulation based on GEANT3 \cite{geant3}. 
Corrections are done to primary charged particle correlations and multiplicities.
Correction factors obtained with these two generators 
are found to agree within 1\% and the difference is neglected.
Three independent correction procedures are investigated.

In the first procedure, the correction factors for $\bcor$ are obtained as the ratio of $\bcor$ 
obtained at generator level (true value)
to $\bcor$  after detector response simulation (measured value).
In the second procedure the correction factors are obtained for $\av{\nB\nF}$, $\av{\nB}$, $\av{\nF}$ and $\av{n_{\rm F}^2}$ separately 
and $\bcor$ is obtained from the corrected moments.
The third procedure takes into account approximately linear dependence of $\bcor$  on $\av{\nF}$ 
when $\av{\nF}$  varies  with cuts,
and each corrected value of $\bcor$ is found by extrapolation to the corrected value of~$\av{\nF}$. 

\begin{table}[t]
\centering
\begin{tabular}{r|c|c|c}
\hline %
Error source  &  $0.9$ TeV&  $2.76$ TeV&  $7$ TeV\\
\hline 
Number of TPC space-points&  0.5--3.0 &    0--0.1&  0.2--0.7 $$  \\
Number of  ITS space-points &  0.6--1.9&  --&0.2--1.4$$ \\
DCA & 3.0--4.0 & 1.0--1.8 &0.1--1.0	$$  \\
Vertex position along the beam line
& 0.2--1.1 &0--1.0	&0--0.7 $$  \\
$\bcor$ correction procedure  &  2.5--4.0&  2.2--4.2	&	1.6--2.8 $$  \\
Event pile-up &  $<$1& $<$1 	&	$<$1 $$  \\
\hline
Total ($\%$) &  3.4--4.5&  2.8--4.2	&	2.0--3.0 $$  \\
\hline
\end{tabular}
\caption{Sources of systematic errors of $\bcor$ measurements
in $\eta$-windows of  width $\delta\eta=0.2$,  and their contributions (in $\%$).
The minimal and maximal estimated values are indicated for each given source. }
\label{tab:err}
\end{table}

It was found that results of all three procedures agree within 1.6-4.2\% (see Table~\ref{tab:err}),
thus proving the robustness of $\bcor$ determination.
The second procedure was chosen 
as the most direct and  commonly used to produce the final corrected value of $\bcor$.
Correction factors increase the values of $\bcor$, obtained for standard cuts, by 
6-10\,$\%$  for analysis in $\eta$-windows and
9-18\,$\%$ for  analysis in $\eta$-$\phi$  windows and in $\pt$ intervals.
By varying the selection cuts (vertex- , DCA-  and track selection cuts), correction procedures, 
and by comparison of the high and low pile-up runs,
the systematic uncertainties on  $\bcor$
have been estimated.
Adding all contributions in quadrature, 
the total systematic uncertainties  
are below 4.5\% (4.2\%, 3\%) at  $\sqrt{s}=0.9$ (2.76, 7)~TeV
for the $\bcor$ analysis in $\eta$-separated windows, and 
6\% for analysis in $\eta$-$\phi$ separated windows at  $\sqrt{s}=0.9$ and  7~TeV.
For the  $\bcor$ analysis in $\pt$ intervals for 7~TeV, the systematic uncertainties are less than 8\%.
Statistical errors  are small and within the symbol sizes for data points in the figures.
A summary of the contributions of systematic uncertainties for $\bcor$  in $\eta$-separated windows
with the width $\delta\eta=0.2$ is presented  in  Table~\ref{tab:err}.

\section{Multiplicity correlations in windows separated in pseudorapidity}
\label{sec:correlationsInEtaWithFullPhi}

\begin{figure}[b]
\centering
\resizebox{1.00\textwidth}{!}{%
\begin{overpic} [trim=0.cm 0cm 0.3cm 0.2cm, clip=true]{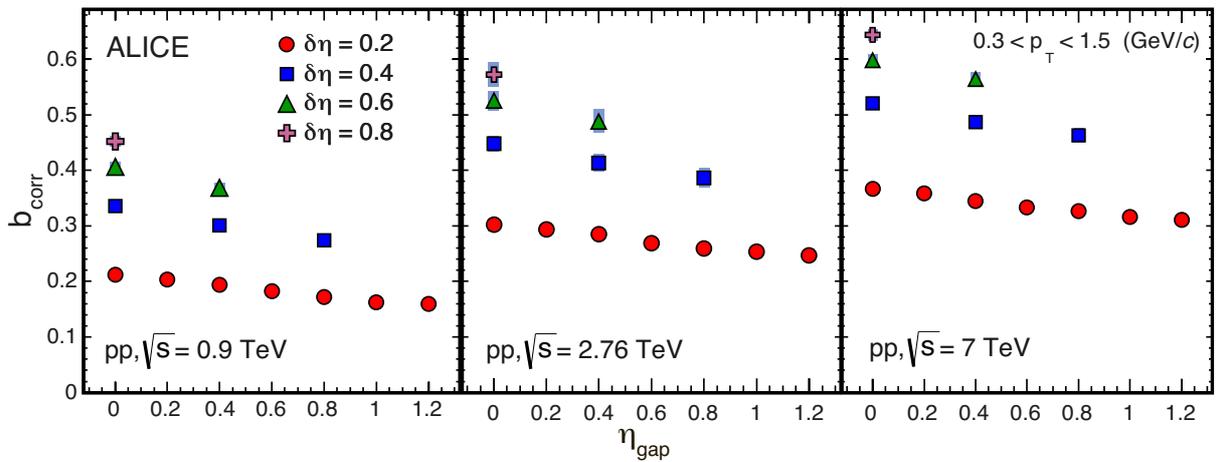} 
\end{overpic}
}
\caption{Forward-backward correlation strength $\bcor$ as function of $\eg$  
and for different windows widths  $\delta\eta$=0.2, 0.4, 0.6 and 0.8 in pp collisions  
at $\sqrt{s}=0.9$, 2.76 and 7~TeV.
}
\label{fig:fullPhiAllWindows}      
\end{figure}

\begin{figure}
\centering
\resizebox{1.00\textwidth}{!}{
\begin{overpic}[trim=0.0cm 0.0cm 0.0cm 0.cm, clip=true]{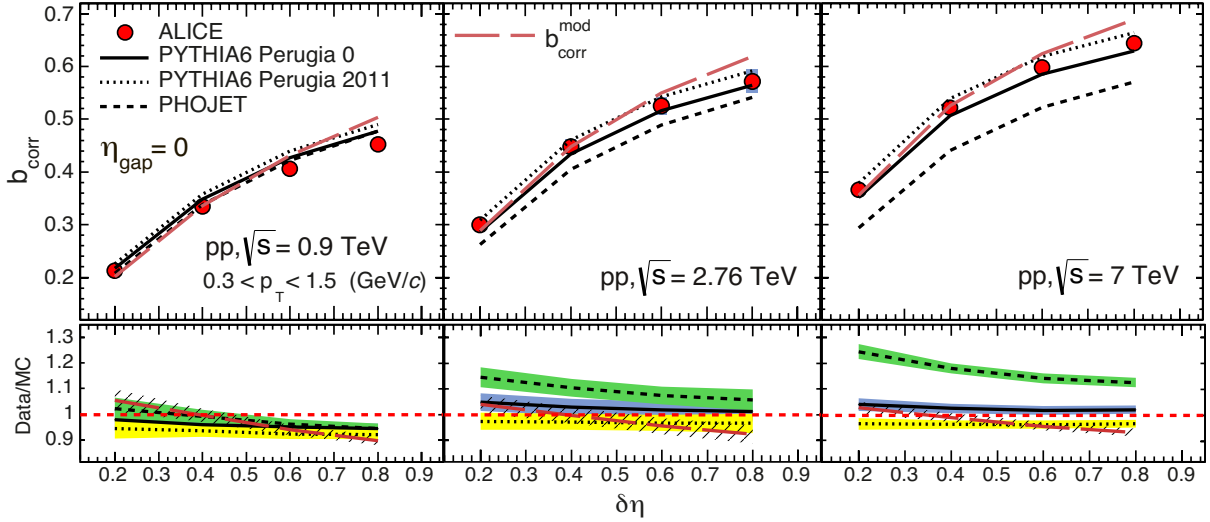} 
\end{overpic}
  }
\caption{ Correlation strength $\bcor$  as a function of $\delta\eta$ for $\eg=0$  in pp collisions 
for   $\sqrt{s}=0.9$,   2.76 and  7~TeV.
The MC results from PYTHIA6 Perugia 0 (solid line), Perugia 2011 (dotted line) and PHOJET
(dashed line), 
calculated at generator level,
are shown for comparison. The bottom panels  show the ratio of $\bcor$ between data and MC.
The red dashed curves correspond to the model of independent particle emission from a fluctuating source (see text).
}
\label{fig:bCorrFromEtaWindowsWidth}     
\end{figure}

\subsection{Dependence on the gap between windows}
Fig.\,\ref{fig:fullPhiAllWindows}  shows the FB multiplicity correlation coefficient $\bcor$
as a function of $\eg$ and 
for different widths of the $\eta$ windows ($\delta\eta$)  
in pp collisions at  the three collision energies.  
For each $\sqrt{s}$, $\bcor$  
is found to decrease slowly with  increasing  $\eg$, 
while maintaining a substantial pedestal value throughout the full $\eg$ range.

\subsection{Dependence on the width of windows}
\label{sec:depOnWindowWidth}
The  $\delta\eta$-dependence for adjacent ($\eg = 0$),  symmetrical windows with respect to $\eta=0$
is shown in Fig.~\ref{fig:bCorrFromEtaWindowsWidth}.
For all collision energies, the correlation coefficient increases non-linearly with $\deta$.
This trend is quite well described by PYTHIA6 and PHOJET, although the agreement  worsens  
with increasing $\sqrt{s}$.
This  $\delta\eta$-dependence can be  
understood, along with other approaches 
\cite{CapellaKrzywicki,BKPV04, VV10}, 
in a simple model with  event-by-event multiplicity fluctuations
and  random distribution of produced particles in pseudorapidity.
In this model, the multiplicity in an $\eta$ interval containing the fraction $p$ 
of the mean multiplicity $\av{N}$ in the full
$\eta$-acceptance  is binomially distributed and its mean square is given by

\beq\label{N2}
\av{n_{\rm F}^2} = \av{n_{\rm B}^2}= p(1-p) \av{N} + p^2  \av{N^2}  \ ,
\eeq
where $N$ is the charged particle multiplicity measured in the pseudorapidity interval $Y$ and 
\beq\label{p}
p=\frac{\av{\nF}}{\av{N}}=\frac{\av{\nB}}{\av{N}}=\frac{\delta\eta}{Y}
  \ .
\eeq
One can connect the multiplicity fluctuations
in the full $\eta$-acceptance considered in this analysis
($Y=1.6$) with the correlation strength $b_{\rm corr}$ (see Appendix \ref{app:formulae}):
\beq\label{bcorrDy}
b^{\rm mod}_{\rm corr}=\frac{\alpha\Dy/Y}{1+\alpha\Dy/Y}   \ ,
\eeq
where
\beq\label{a}
\alpha=\frac{\sigN}{\av{N}}-1  \ .
\eeq

Note that using Eq.\ \ref{N2} and Eq.\ \ref{p} one can write the Eq.\ \ref{bcorrDy} also in the following form:
\beq\label{bcorrnF}
b^{\rm mod}_{\rm corr}=1-\frac{\av{\nF}}{\sigma^2_\nF}  \ .
\eeq

From the measured ratio of the multiplicity variance $\sigN\equiv\av{N^2}-\av{N}^2$
in $Y=1.6$  
to the mean value $\av{N}$ 
we obtain the value of $\alpha$
at  $\sqrt{s}=0.9$, 2.76 and 7~TeV 
to be 2.03, 3.25
and 4.42, respectively, with a systematic  uncertainty of about 5\%. 
The $b^{\rm mod}_{\rm corr}(\delta\eta)$-dependences calculated by Eq.\ \ref{bcorrDy} 
are shown in Fig.\,\ref{fig:bCorrFromEtaWindowsWidth} as red dashed lines.
At $\eg = 0$ the $b_{\rm corr}$($\delta\eta$) dependence 
is well described by this simple model.
However, this  model is not able to describe the dependence of $\bcor$ on $\eg$  
in Fig.\,\ref{fig:fullPhiAllWindows} because it  does not take into account the SRC contribution  mentioned above. 

\subsection{Dependence on the collision energy}
Figure \ref{fig:fullPhiAllWindows} shows that the pedestal value of $\bcor$ increases 
with $\sqrt{s}$, 
while the slope of the $\bcor(\eg)$ dependence  stays approximately constant.
This indicates that the contribution of the short-range correlations has a very weak $\sqrt{s}$-dependence, while the long-range multiplicity correlations  play a dominant role in pp collisions 
and their strength increases significantly  with $\sqrt{s}$. 
Note that this increase cannot be explained
by the increase of the mean multiplicity alone. If, at different energies, we choose window sizes such that the mean multiplicity stays constant
the increase is still observed (see Table~\ref{tab:energyDep}).

In the framework of the simple model described by Eq.\ \ref{bcorrDy} and \ref{a} the increase of the correlation coefficient corresponds to the increase of the event-by-event multiplicity fluctuations with $\sqrt{s}$ characterized by the ratio $\sigma^2_N/\av{N}$. 

A strong energy dependence and rather large $\bcor$ values were previously reported by the UA5 collaboration~\cite{ua5} and 
recently by the ATLAS Collaboration \cite{atlas-lrc}. 
However, as we see in Fig.\,\ref{fig:bCorrFromEtaWindowsWidth}, the correlation coefficient depends in a non-linear way on the width of the pseudorapidity window.  
One has to take this fact into account when comparing
the correlation strengths obtained
under different experimental conditions.
In particular, it explains the small values of  $\bcor$  observed by the STAR collaboration at RHIC (pp, $\sqrt{s}= 200$~GeV) \cite{STAR-FBC}, 
where  narrow FB windows ($\delta \eta = 0.2$) were considered, while in previous pp and $\rm p\overline{\rm p}$  
experiments wider windows of a few units of pseudorapidity were used.

\begin{table}[t!f]
\centering
\begin{tabular}{c|c|c|c|c}
\hline
$\sqrt{s}$ (TeV) &  window width $\delta\eta$ & $\av{n_F}$  &  $\bcor$ ($\eg$=0) & $\bcor$ (max. $\eg$)\\ \hline
$0.9$ & 0.54 & 1.17 &  $0.39 \pm 0.01$ &           $0.35 \pm 0.01$           \\ \hline
$2.76$ & 0.4 & 1.17 &  $0.44 \pm	0.02$ &         $0.38 \pm 0.02$     \\ \hline
$7$ & 0.33  & 1.17 &  $0.48 \pm 0.01$  &        $0.43 \pm 0.01$       \\ \hline
\end{tabular}
\caption{ Correlation strength $\bcor$ in pp collisions at   $\sqrt{s}=0.9$, 2.76 and 7~TeV in  windows with equal mean multiplicity $\av{n_{\rm F}}$ 
and the corresponding values of $\deta$.
Values  are shown for 
adjacent  windows ($\eg$=0) and for  windows with maximal  $\eg$ 
within $|\eta| < 0.8$.
The uncertainty on $\av{n_F}$  is about 0.001.}
 \label{tab:energyDep}
\end{table}

\begin{figure}[t!b]
\resizebox{1.00\textwidth}{!}{
\begin{overpic}[scale=1.0, clip=true, angle=0,trim=0.0cm 0cm 0.0cm 0.0cm]{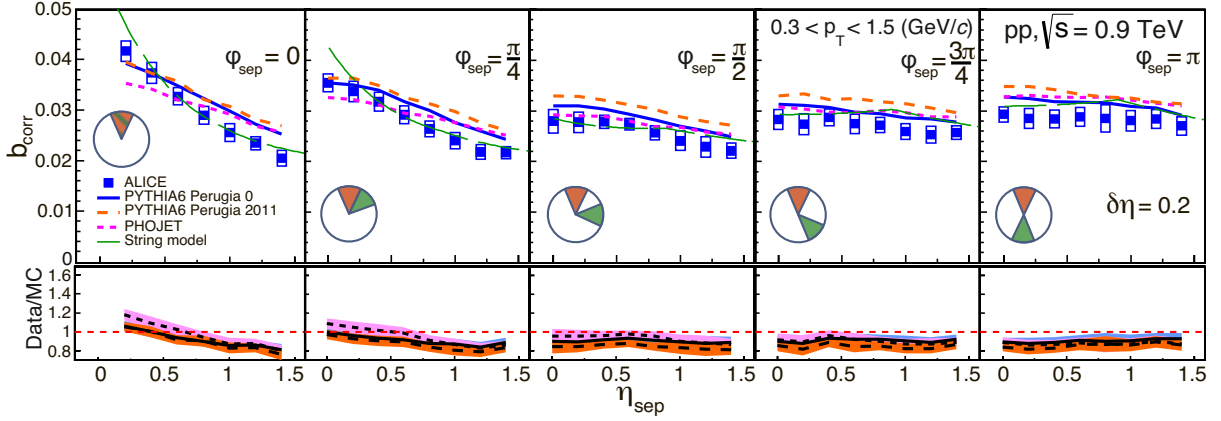}
\end{overpic}
}
\caption{
Correlation strength 
$\bcor$ for separated $\eta$-$\varphi$ window pairs at
 $\sqrt{s}=0.9$ TeV as a function of $\eta$
separation, with fixed window width $\delta\eta$=0.2 and $\delta\varphi$=$\pi$/4.
The panels are for different separation distances 
between the two azimuthal sectors: 
$\varphi_{\rm sep}=0$, $\pi/4$,  $\pi/2$,  $(3/4)\pi$ and  $\pi$.
MC results from  PYTHIA6 Perugia 0 (blue lines), Perugia 2011 (orange dashed lines) and PHOJET
(pink dashed lines)
and string model \cite{Vechernin:2012bz} 
(thin green lines) are also shown. 
The bottom panels  
show the ratio $\bcor$ between  data and MC results. }
\label{fig:pp_900GeV_pads_8by8_5pads}       
\end{figure}

\begin{figure}
\resizebox{1.00\textwidth}{!}{
\begin{overpic}[scale=1.0, clip=true, angle=0,trim=0.0cm 0cm 0.0cm 0.0cm]{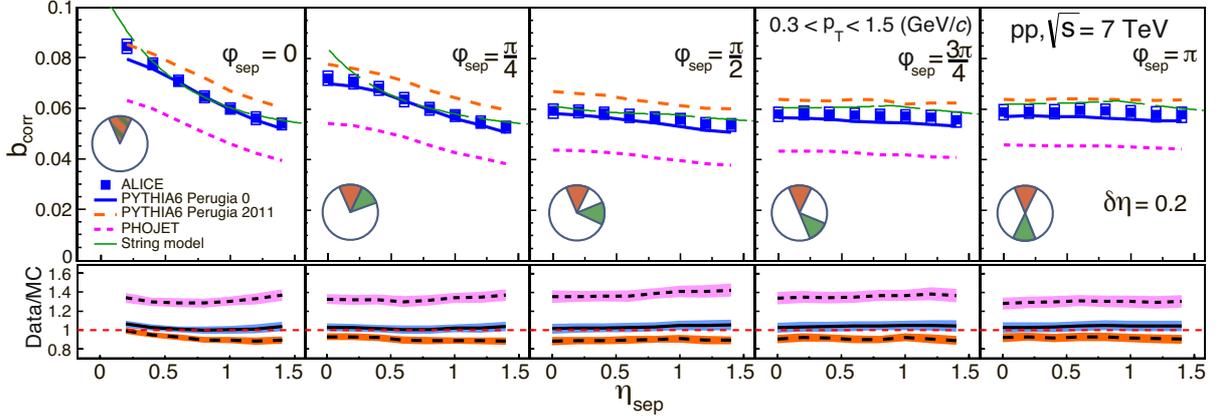}
\end{overpic}
}
\caption{
Correlation strength $\bcor$ for separated $\eta$-$\varphi$ window pairs at
$\sqrt{s}=7$ TeV as a function of $\eta$ separation. The legend is the same as for Fig.\,\ref{fig:pp_900GeV_pads_8by8_5pads}.  }
\label{fig:pp_7TeV_pads_8by8_5pads}   
\end{figure}

\section{Multiplicity correlations in windows separated in pseudorapidity and azimuth}
\label{sec:azimuthalExperimentalResults}

Multiplicity correlations are also studied in different configurations of forward and backward azimuthal sectors. 
These sectors are chosen in separated forward and backward pseudorapidity windows of width $\deta=0.2$ and
 $\delta \varphi = \pi/4$  as shown in Fig.~\ref{fig:windowsPhiSectors},
resulting in 5 pairs with different $\varphi$-separation.

Figs. \ref{fig:pp_900GeV_pads_8by8_5pads} 
and \ref{fig:pp_7TeV_pads_8by8_5pads} 
show the azimuthal dependence 
of $\bcor$ as a function of different $\esep$, for 0.9 and 7~TeV, respectively.
Data are compared to PYTHIA6 (tunes Perugia 0 and Perugia 2011), PHOJET
and a parametric string model \cite{Vechernin:2012bz}. 
 
The string model fitted to our data helps to understand in a simple way
the origins of the $\bcor$ behaviour. 
There are two contributions to $\bcor$ in this model.
The short-range (SR) contribution   
originating from the correlation between particles
produced from the decay of a single string 
and the long-range (LR) contribution   
arising from event-by-event fluctuations of the number of strings.
The energy dependence of the fitted parameters demonstrates that
SR parameters stay constant with $\sqrt{s}$ while the normalized variance of the number of strings,  
the only LR parameter of the model, increases by a factor of three.

\begin{figure}[h]
\subfigure[Subfigure 1 list of figures text][]
{
 \begin{overpic}[width=0.48\textwidth]{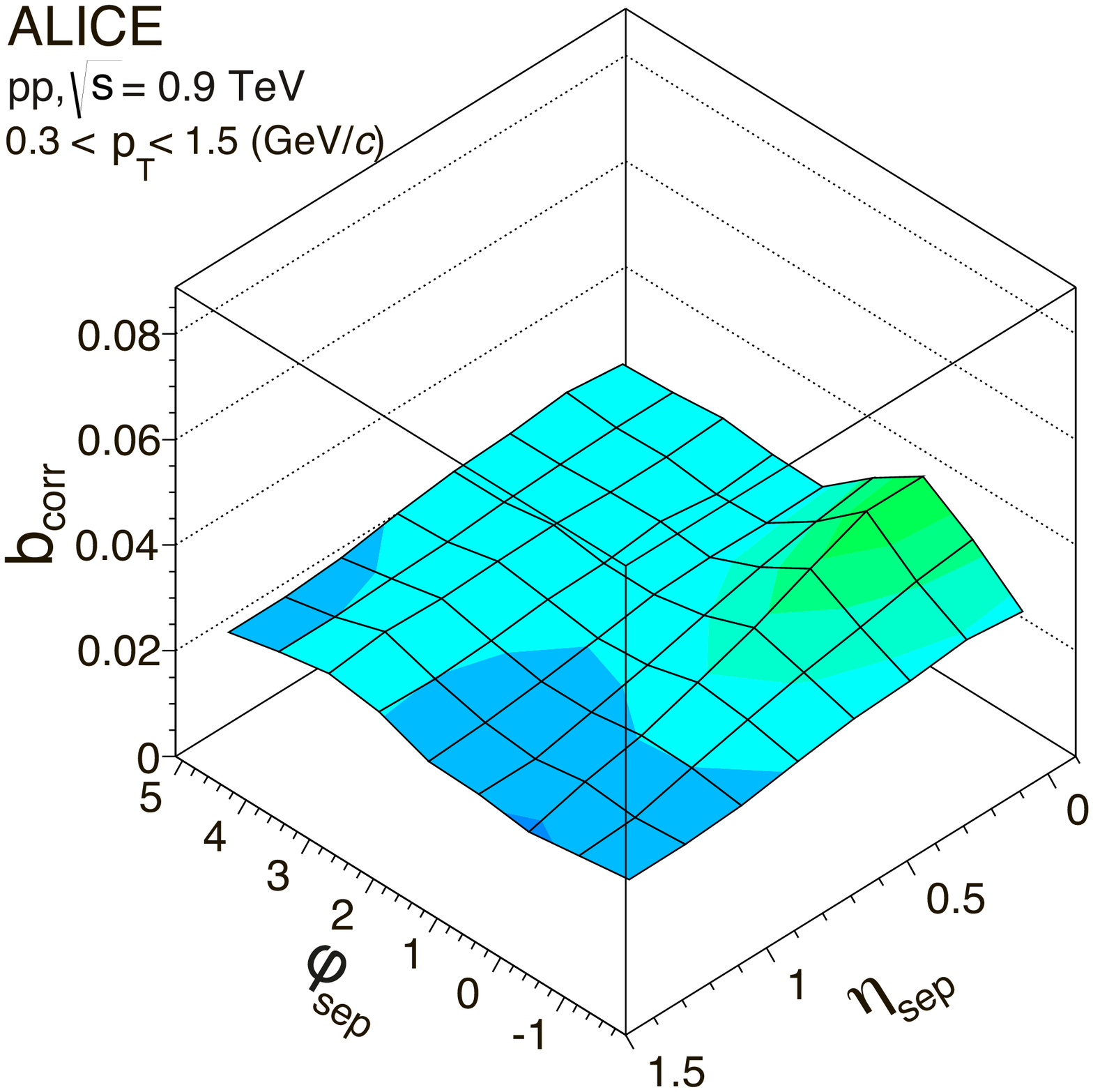}
 \end{overpic}
\label{fig:pp_900GeV_3D_8by8}
}
\hfill
\subfigure[Subfigure 2 list of figures text][]
{
  \begin{overpic}[width=0.48\textwidth]{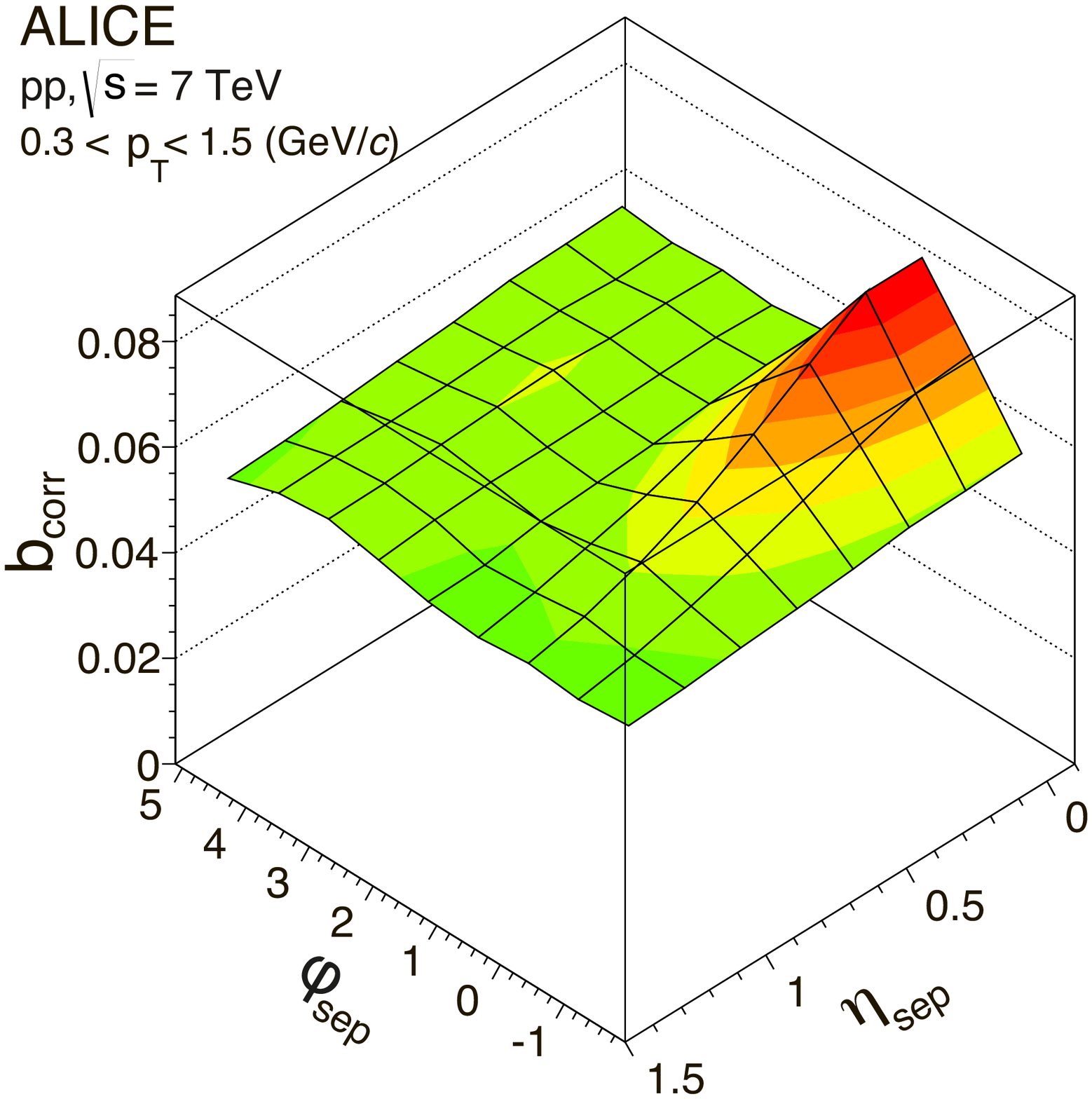}
    \end{overpic}
\label{fig:pp_7TeV_3D_8by8}
}
\caption{2D representation of  $\bcor$ 
at  (a)  $\sqrt{s}=0.9$~TeV 
and 
(b) at  $\sqrt{s}=7$~TeV
for separated $\eta$-$\varphi$  
window pairs with $\delta\eta$=0.2 and $\delta\varphi$=$\pi$/4.
To improve visibility, the point ($\esep$,$\fsep$)=(0,0) and thus $\bcor$=1 is limited 
to the level of the maximum value in adjacent bins.
}
\label{fig:pp_3DbothEnergies_8by8}
\end{figure}

The 2-dimensional distribution of  $\bcor$ as a function of 
$\esep$ and $\fsep$ 
is shown in Fig.\,\ref{fig:pp_3DbothEnergies_8by8} 
for   $\sqrt{s}=0.9$ and 7~TeV.
The qualitative behaviour of $b_{\rm corr}$  
resembles the results obtained for two-particle angular correlations: 
near-side peak and recoil away-side structure.  
The connection between the FB correlation and two-particle correlation function
is discussed in detail in \cite{ Vechernin:2012bz, Vechernin:2013vpa,CapellaKrzywicki,Voloshin02}.

The shapes of the correlation functions clearly indicate 
two contributions to the forward-backward
multiplicity correlation coefficient.
The SR contribution is concentrated within a rather limited region in the $\eta$-$\varphi$ plane
within one unit of pseudorapidity
and $\pi$/2 in azimuth,  
while the LR contribution  manifests itself as a common pedestal in the whole region of observation.

The strength of multiplicity correlations
measured in $\eta$ and $\eta$-$\varphi$ windows
is compared to the results obtained with PYTHIA6 \cite{pythia} (tunes Perugia 0 and Perugia 2011)
and PHOJET \cite{phojet,phojet2} Monte Carlo generators (MC).
The detailed overview of key features of these generators can be found
in \cite{MCnet}.
Recent Perugia tunes for PYTHIA6 are described in \cite{perugiaTunes}.

\begin{figure} [!t]
\resizebox{0.98\textwidth}{!}{
\begin{overpic}{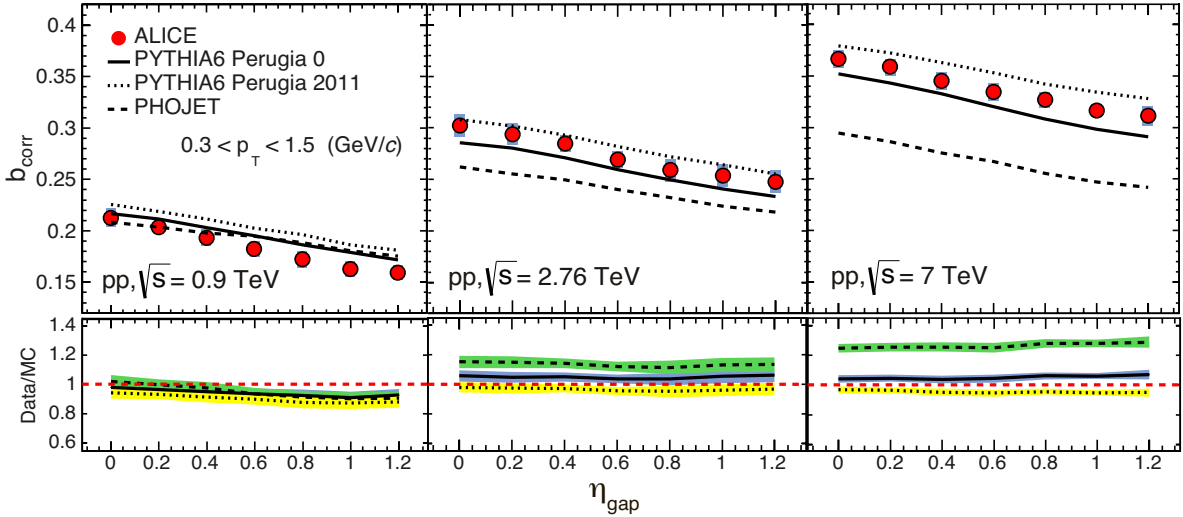} 
\end{overpic}
  }
\caption{ 
Correlation strength 
$\bcor$ as a function of $\eta_{\rm gap}$   in pp collisions for
data  taken from Fig.\,\ref{fig:fullPhiAllWindows}
and compared to MC generators
 PYTHIA Perugia 0 (solid line), Perugia 2011 (dotted line) and PHOJET
(dashed line)  for  $\sqrt{s}=0.9$,  2.76 and   7~TeV collision energies,
windows width $\deta$ is 0.2. The bottom panels 
show the ratio of the data to MC.}
\label{fig:bCorrFromEtaGap}      
\end{figure}

In Fig.\,\ref{fig:bCorrFromEtaGap} the comparison of $\bcor$
as a function of $\eta_{\rm gap}$ for $\deta=0.2$ at  $\sqrt{s}=0.9$, 2.76 and 7~TeV
with the results obtained with different MC generators
is shown. All models describe the data at  $\sqrt{s}=0.9$~TeV reasonably well,
while larger discrepancies are observed at 2.76 and 7~TeV, with
PYTHIA giving a better description of the data than PHOJET.
Qualitatively similar  conclusions
can be drawn from the comparison of  the $\deta$-dependence in experimental data and MC as shown in 
Fig.~\ref{fig:bCorrFromEtaWindowsWidth}.

\begin{figure*} 
\centering
\resizebox{0.67\textwidth}{!}{%
\includegraphics[trim=0.2cm 0.1cm 0.7cm 0.3cm, clip=true]{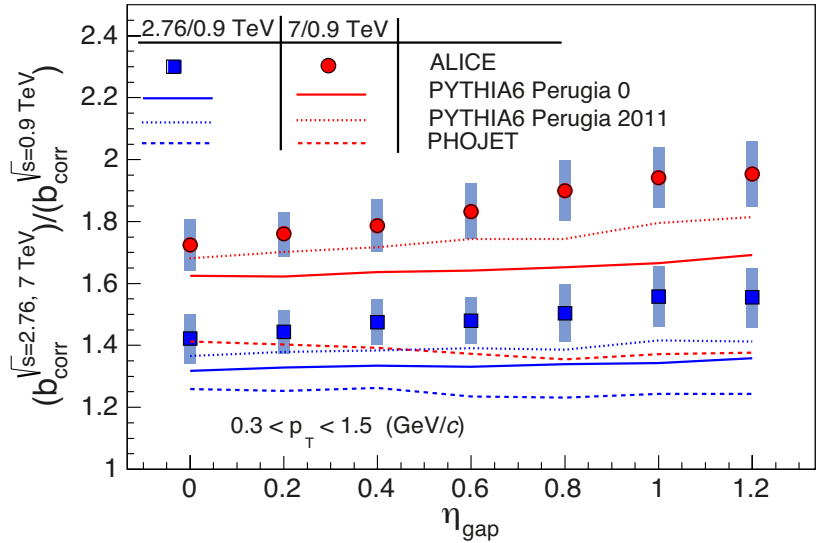}
  }
\caption{Ratio of $\bcor$ at 2.76 (blue squares) and 7~TeV (red circles) with respect to 0.9~TeV vs. $\eta_{\rm gap}$.
The calculations from MC generators are also shown:
PYTHIA Perugia 0 (solid line), Perugia 2011 (dotted line) and PHOJET
(dashed line),
 for 2.76 TeV (blue lines) and 7 TeV (red lines).  }
\label{fig:bRatioWrt900GeV}  
\end{figure*}

Note that PYTHIA also describes the correlations in $\eta$-$\varphi$ windows reasonably well,
see Figures \ref{fig:pp_900GeV_pads_8by8_5pads} and \ref{fig:pp_7TeV_pads_8by8_5pads},
while PHOJET gives a good description only for $\sqrt{s}=0.9$~TeV
and significantly underestimates the data at 7~TeV.

The difference between the experimental data and
the results obtained with MC generators
is more visible in
Fig.\,\ref{fig:bRatioWrt900GeV}, which compares the measured ratio of $\bcor$
at $\sqrt{s} = 2.76$ and 7~TeV with respect to 0.9~TeV
as a function of $\eg$ to MC calculations.
The  measured ratios show an increasing trend as a function of $\eta_{\rm gap}$,
while PYTHIA and PHOJET underestimate the ratios
and exhibit a flatter $\eta_{\rm gap}$ dependence.

It is important to note that, in the framework of PYTHIA, the observed LR part  of $\bcor$ 
(the pedestal in Fig. \ref{fig:pp_3DbothEnergies_8by8})
is dominated by  multiple parton-parton interactions (MPI).
This supports earlier results \cite{PS},
in which the FB correlations in pp collisions were studied
by MC simulations with recent tunes of the PYTHIA6 at $\sqrt{s}=0.9$~TeV.
Hence, the observed dependence of $\bcor$ on collision energy and on different configurations of rapidity and azimuthal windows 
adds new constraints on phenomenological models for multi-particle production.

\section{Dependence of FB multiplicity correlation strength on the choice of $\pt$ intervals}
\label{sec:ptBinsStudy}

The  behaviour of FB multiplicity correlation strength  was also studied as  a function of
$\pt$ of registered  particles. 
These studies were motivated  by a recent paper 
by the ATLAS collaboration \cite{atlas-lrc},
which reported a decrease in the multiplicity correlation strength 
with increasing $\ptmin$. 
However, as we have observed
in Section~\ref{sec:depOnWindowWidth},
there is a strong non-linear dependence of $b_{\rm corr}$ on the size of pseudorapidity windows
and, hence, on the mean multiplicity $\av{n_{\rm ch}}$ in the window
(see Eq.\ \ref{p}, \ref{bcorrDy}, and Fig.\,\ref{fig:bCorrFromEtaWindowsWidth}).
In order to demonstrate that the strong $\ptmin$ dependence is not a trivial multiplicity dependence, 
in our analysis we use $\pt$ intervals with the same  $\av{n_{\rm ch}}$.
To this end, the correlation strength $\bcor$ is studied for
five $\pt$ intervals within $0.3<\pt<6$~GeV/$c$ 
at $\sqrt{s}=7$\,TeV:  
0.3--0.4,
0.4--0.52,
0.52--0.7,
0.7--1.03 and
1.03--6.0 (GeV/c). 
In each $\pt$ interval,  the corrected mean multiplicity 
$\av{\nF}=0.157$ with a systematic uncertainty about 2\%.
Correlations are studied in $\eta$ and $\eta$-$\varphi$ FB-windows configurations.
Note that in case of windows chosen symmetrically with respect to $\eta = 0$ 
the definition of $\bcor$ given by \eqref{b-alt} coincides
with the correlation coefficient $\rho^n_{\rm FB}$ used  in the ATLAS analysis.

\begin{figure}
\begin{center}
\begin{overpic}[width=0.53\textwidth]{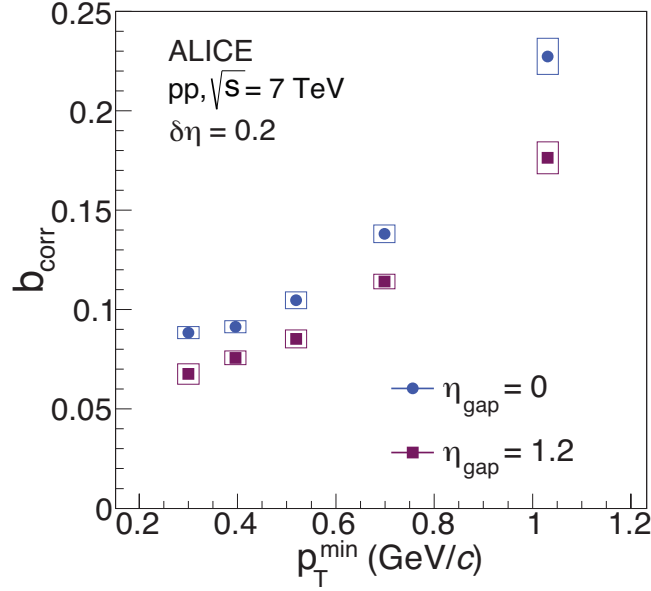}
\end{overpic}
\end{center}
\caption{
Correlation strength $\bcor$ at $\sqrt{s}=7$~TeV for separated pseudorapidity window pairs,
 measured in $\pt$ intervals with same $\av{n_{\rm ch}}$,
as a function of $\ptmin$ for each interval.
Values are shown for  $\eg$=0, 1.2
with  $\delta\eta$=0.2. 
}
\label{fig:ptBins_FROM_PT_fullPhi}
\end{figure}

\begin{figure*}
\centering
$\begin{array}{ccc}
\begin{overpic}[width=0.46\textwidth]{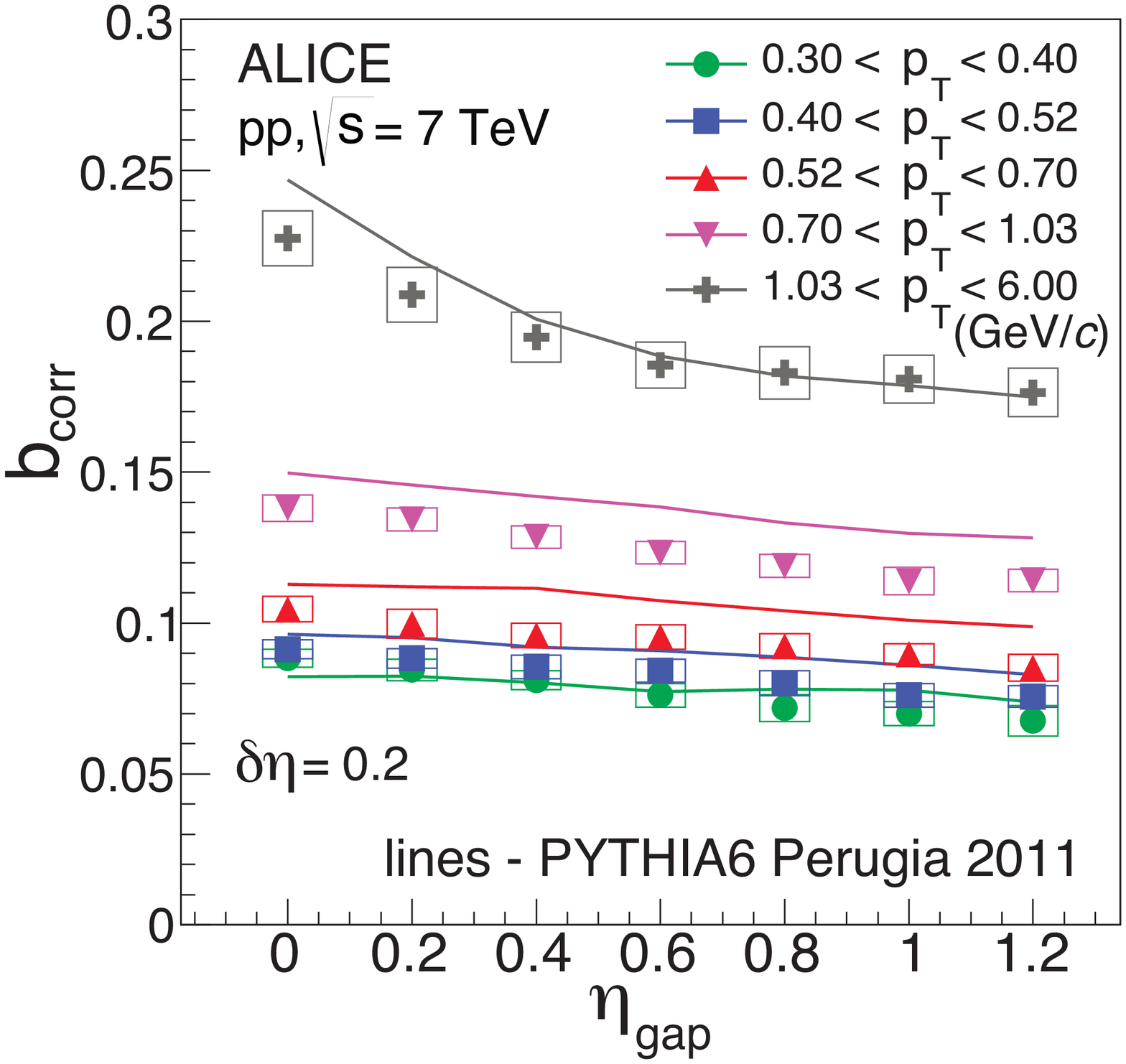}
\end{overpic}
\label{fig:ptBins_MC_Pythia_fullPhi}
&
\begin{overpic}[width=0.46\textwidth]{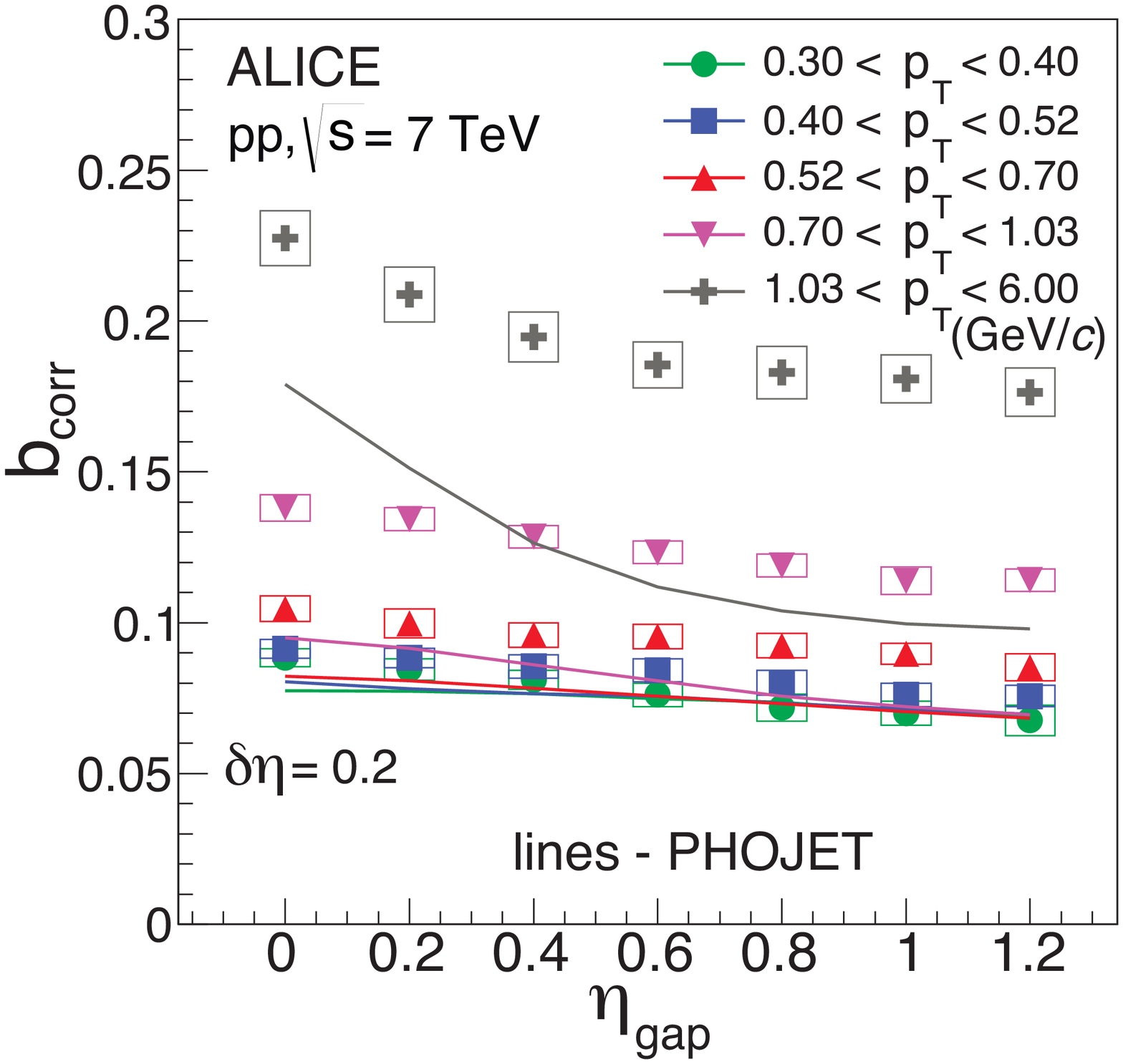}    
\end{overpic}
\end{array}$
\caption{
Correlation strength $\bcor$ at $\sqrt{s}=7$~TeV for separated pseudorapidity window pairs,
measured in $\pt$ intervals with same $\av{n_{\rm ch}}$
as a function of $\eg$. Windows of  width $\delta\eta$=0.2.
Left and right panels contain same data points, lines correspond to  PYTHIA6 Perugia 2011 (left) and
to PHOJET (right).
}
\label{fig:ptBins_MCall_fullPhi}
\end{figure*}

Fig.\,\ref{fig:ptBins_FROM_PT_fullPhi} shows $\bcor$ as a function of $\ptmin$ 
for  $\eg$ = 0 and 1.2.
Systematic uncertainties are shown as  rectangles, statistical uncertainties  are negligible.
We find that $\bcor$ increases with $\ptmin$ for both values of $\eta_{\rm gap}$, in contrast to the results reported in \cite{atlas-lrc}.
This result can be understood if one takes into account
that the multiplicity fluctuations in a given window
are closely connected with the two-particle correlation strength \cite{CapellaKrzywicki,Voloshin02}.
In the simple model with the event-by-event multiplicity fluctuations
and random distribution of produced particles in pseudorapidity,
discussed in Section \ref{sec:depOnWindowWidth},
 Eq.\ \ref{bcorrnF} allows us to discuss the observed dependence of the correlation coefficient $\bcor$
on the $\pt$-binnings for the case of $\eta_{\rm gap} = 0$ (Fig.\,\ref{fig:ptBins_FROM_PT_fullPhi}).
One sees that the imposed condition $\av{\nF}$=$const$
eliminates  the dependence of  $\bcor$  on the multiplicity.
The ratio $1/\sigma^2_\nF$ decreases and $b^{\rm mod}_{\rm corr}$ increases with increasing  $\ptmin$.

As mentioned above, in the approach used in \cite{atlas-lrc}
the dependence of the correlation strength on the $\ptmin$
of charged particles was studied without cuts on $\ptmax$, which
leads to a decrease of the correlation  $\bcor$ with increasing $\ptmin$.
This result can also be illustrated with the help of  Eq.\ \ref{bcorrnF}.
In this case
$\av{\nF}$ decreases with increasing  $\ptmin$ and $\av{\nF}/\sigma^2_\nF$ increases
(approaching the Poisson limit $\sigma^2_\nF=\av{\nF}$)
leading to the decrease of $b^{\rm mod}_{\rm corr}$.
Thus, the difference of the results in these two approaches
can be  qualitatively understood using  Eq.\ \ref{bcorrnF}.

Fig. \ref{fig:ptBins_MCall_fullPhi} 
shows  $\bcor$  as function of $\eg$ for different $\pt$ intervals.
Fig.\,\ref{fig:ptBins_MCall_fullPhi}~(a)  compares data to
PYTHIA6 tune Perugia 2011.
The general trend of $\bcor$ increasing  with  higher $\ptmin$ for all $\eg$   is reproduced by this tune,
with small quantitative deviations.
Fig.\,\ref{fig:ptBins_MCall_fullPhi}~(b) shows the same data in comparison to  PHOJET.
This generator does not describe the data well: 
PHOJET  results are almost
independent of $\ptmin$ and only grow significantly
for the $\pt$ range 1.03--6.00 (GeV/$c$).  
Since experimental data was used to determine the $\pt$ intervals with the same mean multiplicity,
the values of mean multiplicities may vary slightly in case of the MC 
samples for the same $\pt$ intervals.
Deviations from the mean value are within 4\% for PYTHIA6 Perugia 0 and 12\% for PHOJET.

\begin{figure*}
\centering
{
\begin{overpic}[width=0.99\textwidth, clip=true, angle=0,trim=0.0cm 0cm 0cm 0.0cm]
{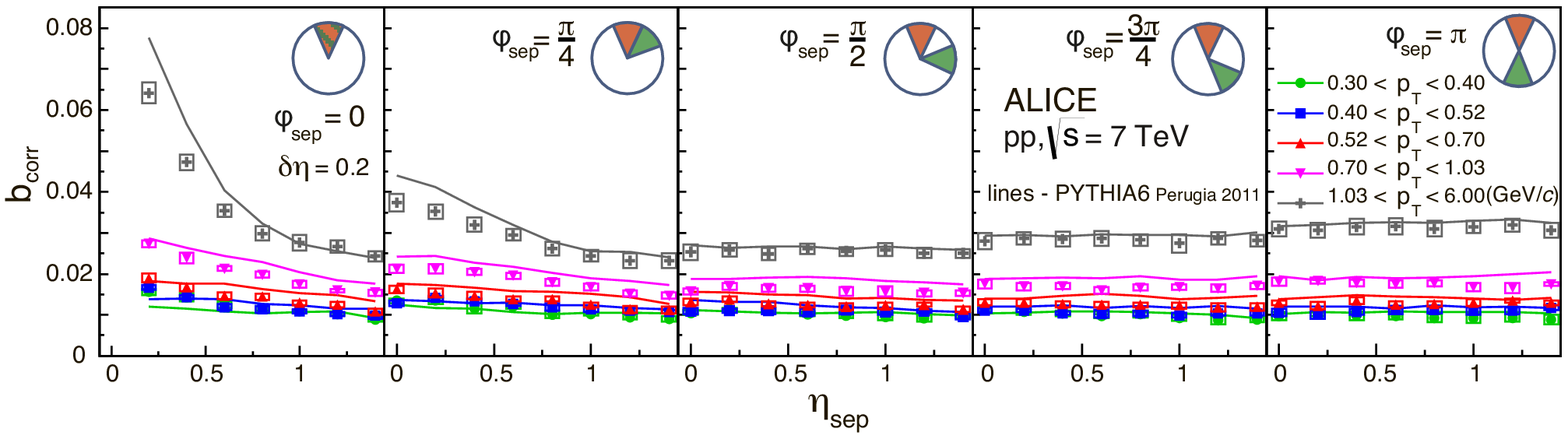} 
\end{overpic}
\label{fig:ptBins_pads_MC_pythia}
}
\\
\vspace{0.3cm}
{
\begin{overpic}[width=0.99\textwidth, clip=true, angle=0,trim=0.0cm 0cm 0cm 0.0cm]
{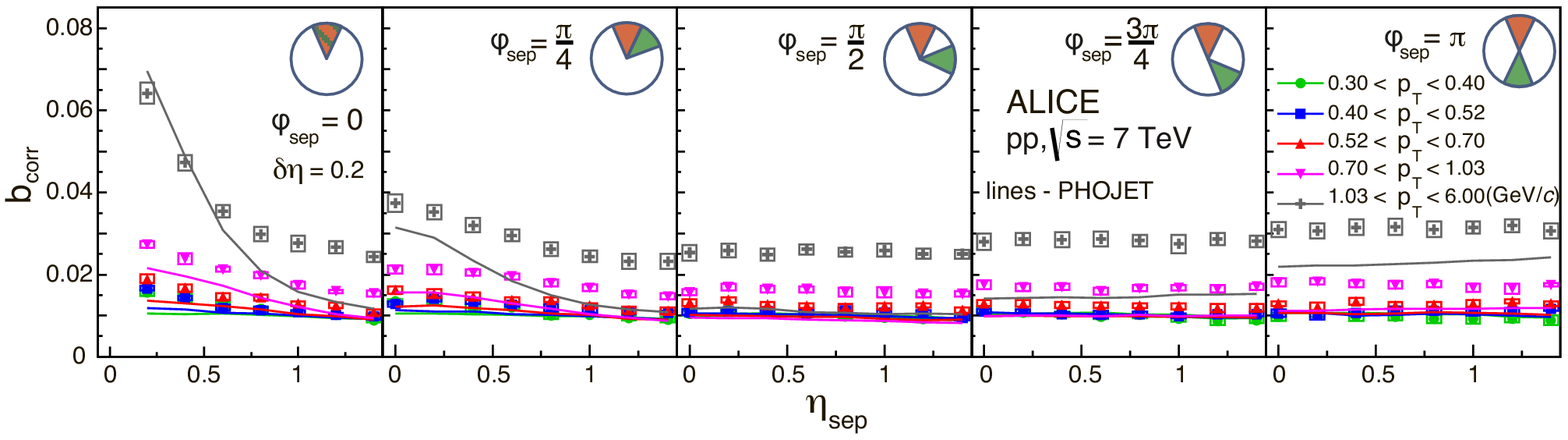}  
\end{overpic}
\label{fig:ptBins_pads_MC_phojet}
}
\caption{Correlation strength $\bcor$ at $\sqrt{s}=7$ TeV
for separated $\eta$-$\varphi$ windows in different $\pt$ intervals with same $\av{n_{\rm ch}}$.
Five $\fsep$ values are shown as a function of $\esep$.
 Windows of  $\delta\eta$=0.2
and $\dphi$=$\pi$/4.
Top and bottom panels and  contain the same experimental data, 
lines correspond to  PYTHIA6 Perugia~2011~(top) and
to PHOJET~(bottom).
}
\label{fig:ptBins_pads_MC_all}
\end{figure*}

The analysis of  $\bcor$   is also performed in
$\eta$-$\varphi$ separated windows in different $\pt$ intervals with the same mean multiplicity 
(for pp collisions at $\sqrt{s}=7$~TeV)
in 8$\times$8 $\eta$-$\varphi$ windows.
Results   are shown in Fig.\,\ref{fig:ptBins_pads_MC_all} and compared to PYTHIA6 and PHOJET calculations.
In addition to the  conclusions that were drawn above from the correlations between $\eta$-separated windows,
some new details are revealed.
In particular, one observes that the PHOJET discrepancy
with the data is especially dramatic at $\fsep$=$\pi$/2, 
where PHOJET shows no dependence of $\bcor$ on the $\pt$ range.
It was shown already in \cite{ALICE:2011ac} that PHOJET has difficulties in description of 
underlying event measurements.

Fig.\,\ref{fig:ptBins_pads_MC_all} shows
that for higher $\pt$ intervals a near-side peak  appears
(see panels for $\varphi_{\rm sep}=0$ and $\pi/4$),
at the same time the $b_{\rm corr}$ in the flat region at $\eta_{sep}> 1$  increases with $\pt$ for all $\varphi_{\rm sep}$ values
(compare panels for $\varphi_{\rm sep}=\pi/2$, $3\pi/4$ and $\pi$).
It should be emphasized that the value of the pedestal (the common constant component in all panels)
increases  with $\pt$.

In near- and away-side azimuthal regions the increase of $\bcor$ with $\ptmin$ can be explained by 
an enhanced number of back-to-back decays and jets. The general rise of $\bcor$ can be related to the increase of the variance 
$\sigN$ in Eq.\ \ref{a},
discussed in the framework of the simple model in Section~\ref{sec:depOnWindowWidth}.

\section{Conclusion}
\label{sec:conclusion}

The strengths of forward-backward (FB) multiplicity correlations  have been measured
in minimum bias pp collisions at $\sqrt{s}=0.9$, 2.76 and 7~TeV using multiplicities 
determined in two separated pseudorapidity windows 
separated  by a variable gap, $\eg$, of up to 1.2 units.
The dependences of the correlation coefficient $\bcor$  
on the collision energy, the width and the position
of pseudorapidity windows have been investigated.
For the first time, the analysis has been also applied for various configurations of the azimuthal sectors selected within these pseudorapidity windows in events at $\sqrt{s}=0.9$ and 7~TeV.

A considerable increase of the FB correlation strength 
with the growth of the collision energy from $\sqrt{s}=0.9$ to 7 TeV is observed.
It is shown that this cannot be explained by the increase of the mean multiplicity alone.
The correlation strength
grows with the width of pseudorapidity windows, while 
it decreases slightly  
with increasing pseudorapidity gap between the windows.
It is shown that  
there is a strong non-linear dependence of the correlation strength on the width of the pseudorapidity windows and hence on the 
mean multiplicity value.

Measurements of the correlation strength 
for various configurations of azimuthal sectors enable the distinction  
of two contributions: short-range (SR) and  long-range (LR) correlations. 
A weak dependence on the collision energy is observed for the SR component while the LR component
has a strong dependence.
For $\eta$-gaps larger than
one unit of pseudorapidity and $\pi$/2 in azimuth the LR contribution dominates.
This contribution forms a pedestal value 
(the common constant component) 
of $\bcor$ increasing with collision energy.

Moreover, pseudorapidity and pseudorapidity-azimuthal  distributions of $\bcor$ have been obtained 
in pp events at $\sqrt{s}=7$~TeV for various  particle transverse momentum intervals.
It is found that the FB correlation strength  increases with the transverse momentum if $\pt$-intervals  with the same mean multiplicity are chosen.

The measurements have been compared to calculations using the PYTHIA and PHOJET MC event generators.
These generators are able to describe the general trends of  $\bcor$ as a function of 
$\deta$, $\eg$ and $\fsep$   
and its dependence on the collision energy.
In  $\pt$-dependent analysis of  $\bcor$,  PYTHIA
describes data reasonably well, while PHOJET fails to describe 
$\bcor$ in azimuthal sectors.
The observed dependences of $\bcor$ add  new constraints on  phenomenological models. In particular the transition between soft and hard processes in pp collisions can be investigated in detail using the $\pt$ dependence of azimuthal and pseudorapidity distributions of forward-backward multiplicity correlation strength $\bcor$.

\newenvironment{acknowledgement}{\relax}{\relax}
\begin{acknowledgement}
\section*{Acknowledgements}
The ALICE Collaboration would like to thank all its engineers and technicians for their invaluable contributions to the construction of the experiment and the CERN accelerator teams for the outstanding performance of the LHC complex.
The ALICE Collaboration gratefully acknowledges the resources and support provided by all Grid centres and the Worldwide LHC Computing Grid (WLCG) collaboration.
The ALICE Collaboration acknowledges the following funding agencies for their support in building and
running the ALICE detector:
State Committee of Science,  World Federation of Scientists (WFS)
and Swiss Fonds Kidagan, Armenia,
Conselho Nacional de Desenvolvimento Cient\'{\i}fico e Tecnol\'{o}gico (CNPq), Financiadora de Estudos e Projetos (FINEP),
Funda\c{c}\~{a}o de Amparo \`{a} Pesquisa do Estado de S\~{a}o Paulo (FAPESP);
National Natural Science Foundation of China (NSFC), the Chinese Ministry of Education (CMOE)
and the Ministry of Science and Technology of China (MSTC);
Ministry of Education and Youth of the Czech Republic;
Danish Natural Science Research Council, the Carlsberg Foundation and the Danish National Research Foundation;
The European Research Council under the European Community's Seventh Framework Programme;
Helsinki Institute of Physics and the Academy of Finland;
French CNRS-IN2P3, the `Region Pays de Loire', `Region Alsace', `Region Auvergne' and CEA, France;
German Bundesministerium fur Bildung, Wissenschaft, Forschung und Technologie (BMBF) and the Helmholtz Association;
General Secretariat for Research and Technology, Ministry of
Development, Greece;
Hungarian Orszagos Tudomanyos Kutatasi Alappgrammok (OTKA) and National Office for Research and Technology (NKTH);
Department of Atomic Energy and Department of Science and Technology of the Government of India;
Istituto Nazionale di Fisica Nucleare (INFN) and Centro Fermi -
Museo Storico della Fisica e Centro Studi e Ricerche "Enrico
Fermi", Italy;
MEXT Grant-in-Aid for Specially Promoted Research, Ja\-pan;
Joint Institute for Nuclear Research, Dubna;
National Research Foundation of Korea (NRF);
Consejo Nacional de Cienca y Tecnologia (CONACYT), Direccion General de Asuntos del Personal Academico(DGAPA), M\'{e}xico, :Amerique Latine Formation academique – European Commission(ALFA-EC) and the EPLANET Program
(European Particle Physics Latin American Network)
Stichting voor Fundamenteel Onderzoek der Materie (FOM) and the Nederlandse Organisatie voor Wetenschappelijk Onderzoek (NWO), Netherlands;
Research Council of Norway (NFR);
National Science Centre, Poland;
Ministry of National Education/Institute for Atomic Physics and Consiliul Naţional al Cercetării Ştiinţifice - Executive Agency for Higher Education Research Development and Innovation Funding (CNCS-UEFISCDI) - Romania;
Ministry of Education and Science of Russian Federation, Russian
Academy of Sciences, Russian Federal Agency of Atomic Energy,
Russian Federal Agency for Science and Innovations and The Russian
Foundation for Basic Research;
Ministry of Education of Slovakia;
Department of Science and Technology, South Africa;
Centro de Investigaciones Energeticas, Medioambientales y Tecnologicas (CIEMAT), E-Infrastructure shared between Europe and Latin America (EELA), Ministerio de Econom\'{i}a y Competitividad (MINECO) of Spain, Xunta de Galicia (Conseller\'{\i}a de Educaci\'{o}n),
Centro de Aplicaciones Tecnológicas y Desarrollo Nuclear (CEA\-DEN), Cubaenerg\'{\i}a, Cuba, and IAEA (International Atomic Energy Agency);
Swedish Research Council (VR) and Knut $\&$ Alice Wallenberg
Foundation (KAW);
Ukraine Ministry of Education and Science;
United Kingdom Science and Technology Facilities Council (STFC);
The United States Department of Energy, the United States National
Science Foundation, the State of Texas, and the State of Ohio;
Ministry of Science, Education and Sports of Croatia and  Unity through Knowledge Fund, Croatia.
Council of Scientific and Industrial Research (CSIR), New Delhi, India
\end{acknowledgement}

\bibliographystyle{utphys}   

\bibliography{bibliography_BibTeX} 


\appendix
\section{A model with random uniform distribution of produced particles in pseudorapidity}
\label{app:formulae}

\def\nF{{n^{}_{\rm F}}}
\def\nB{{n^{}_{\rm B}}}
\def\sigN{{\sigma^2_{\rm N}}}

In a simple model with event-by-event multiplicity fluctuations and random uniform distribution of produced
particles in pseudorapidity the probability to observe $\nF$ particles in some subinterval $\delta\eta$
from the total number of $N$ charged particles produced in the whole pseudorapidity interval $Y$ is given by
the binomial distribution:
\beq\label{binomial}
P_N(\nF)=C_N^\nF p^\nF (1-p)^{N-\nF}  \  ,
\eeq
with $\av{\nF}^{}_N=pN$ and  $\av{n_F^2}^{}_N=p(1-p)N+p^2 N^2$,
where $p\equiv\Dy/Y$. (We consider the case of symmetric windows $\Dy_{\rm F}=\Dy_{B}=\Dy$.)
Averaging then over events with different values of $N$,
\beq\label{PnF}
P(\nF)=\sum_N P(N) P_N(\nF) \  ,
\eeq
we have
\begin{equation}
\label{avr_nF}
\av{\nF}=\sum_\nF  P(\nF) \nF =\sum_\nF  \sum_N P(N) P_N(\nF) \nF  = \sum_N P(N) pN=p\,\av{N}
\end{equation}
and hence
\beq\label{avr}
p=\frac{\av{\nF}}{\av{N}}=\frac{\av{\nB}}{\av{N}}=\frac{\Dy}{Y} \ .
\eeq
In the same way we find
\beq\label{31}
\av{n^2_{\rm F}}=\av{n^2_{\rm B}}=p(1-p)\av N + p^2 \av {N^2}  \ ,
\eeq
\beq\label{31a}
\av{(\nF+\nB)^2}=2p(1-2p)\av N + (2p)^2 \av {N^2} \ .
\eeq
One can rewrite (\ref{avr})--(\ref{31a}) also as
\beq\label{rob}
\frac{\sigFB-\av{\nF+\nB}}{\av{\nF+\nB}^2}=\frac{\sigF-\av{\nF}}{\av{\nF}^2}=\frac{\sigN-\av{N}}{\av{N}^2}\equiv R_{\rm N}
 \ ,
\eeq
since the so-called robust variance $R_{\rm N}$ is the same
for any subinterval of $Y$ in the case of the independent homogeneous distribution
of the particles along $Y$ \cite{Voloshin02}.

Using the  presentation for the covariance
\beq\label{corr}
\av{\nF\nB} -\av{\nF}\av{\nB}\equiv\frac{1}{2}( \sigFB-\sigF-\sigB)   \ ,
\eeq
we can write for the correlation coefficient  in a model-independent way:
\beq\label{bcorr1}
b_{\rm corr}=\frac{ \sigFB-\sigF-\sigB}{2\sigF}   \ .
\eeq
Then combining  (\ref{rob}) and (\ref{bcorr1})
we find
\beq\label{bcorrR}
b^{\rm mod}_{\rm corr}=\frac{\av{\nF}R_{\rm N}}{1+\av{\nF}R_{\rm N}}   \ .
\eeq
Using (\ref{avr}) we can write (\ref{bcorrR}) also as
\beq\label{bcorrDyApp}
b^{\rm mod}_{\rm corr}=\frac{\alpha\Dy/Y}{1+\alpha\Dy/Y}   \ ,
\eeq
where
\beq\label{aApp}
\alpha=\av{N}R_{\rm N}=\frac{\sigN}{\av{N}}-1
  \ .
\eeq

Substituting the expression
\beq\label{rob1}
R_{\rm N}=\frac{\sigF-\av{\nF}}{\av{\nF}^2}
\eeq
from (\ref{rob}) into (\ref{bcorrR}) one finds another presentation for $b^{\rm mod}_{\rm corr}$\,:
\beq\label{bcorrRaApp}
b^{\rm mod}_{\rm corr}=1-\frac{\av{\nF}}{\sigF}   \ .
\eeq

\newpage


\section{The ALICE Collaboration}
\label{app:collab}



\begingroup
\small
\begin{flushleft}
J.~Adam\Irefn{org40}\And
D.~Adamov\'{a}\Irefn{org83}\And
M.M.~Aggarwal\Irefn{org87}\And
G.~Aglieri Rinella\Irefn{org36}\And
M.~Agnello\Irefn{org111}\And
N.~Agrawal\Irefn{org48}\And
Z.~Ahammed\Irefn{org131}\And
I.~Ahmed\Irefn{org16}\And
S.U.~Ahn\Irefn{org68}\And
I.~Aimo\Irefn{org94}\textsuperscript{,}\Irefn{org111}
% ...
\renewcommand\labelenumi{\textsuperscript{\theenumi}~}

\section*{Affiliation notes}
\renewcommand\theenumi{\roman{enumi}}
\begin{Authlist}
\item \Adef{0}Deceased
\item \Adef{idp5870448}{Also at: University of Kansas, Lawrence, Kansas, United States}
\end{Authlist}

\section*{Collaboration Institutes}
\renewcommand\theenumi{\arabic{enumi}~}
\begin{Authlist}

\item \Idef{org1}A.I. Alikhanyan National Science Laboratory (Yerevan Physics Institute) Foundation, Yerevan, Armenia
\item \Idef{org2}Benem\'{e}rita Universidad Aut\'{o}noma de Puebla, Puebla, Mexico
\item \Idef{org3}Bogolyubov Institute for Theoretical Physics, Kiev, Ukraine
\item \Idef{org4}Bose Institute, Department of Physics and Centre for Astroparticle Physics and Space Science (CAPSS), Kolkata, India
\item \Idef{org5}Budker Institute for Nuclear Physics, Novosibirsk, Russia
\item \Idef{org6}California Polytechnic State University, San Luis Obispo, California, United States
\item \Idef{org7}Central China Normal University, Wuhan, China
\item \Idef{org8}Centre de Calcul de l'IN2P3, Villeurbanne, France
\item \Idef{org9}Centro de Aplicaciones Tecnol\'{o}gicas y Desarrollo Nuclear (CEADEN), Havana, Cuba
\item \Idef{org10}Centro de Investigaciones Energ\'{e}ticas Medioambientales y Tecnol\'{o}gicas (CIEMAT), Madrid, Spain
\item \Idef{org11}Centro de Investigaci\'{o}n y de Estudios Avanzados (CINVESTAV), Mexico City and M\'{e}rida, Mexico
\item \Idef{org12}Centro Fermi - Museo Storico della Fisica e Centro Studi e Ricerche ``Enrico Fermi'', Rome, Italy
\item \Idef{org13}Chicago State University, Chicago, Illinois, USA
\item \Idef{org14}China Institute of Atomic Energy, Beijing, China
\item \Idef{org15}Commissariat \`{a} l'Energie Atomique, IRFU, Saclay, France
\item \Idef{org16}COMSATS Institute of Information Technology (CIIT), Islamabad, Pakistan
\item \Idef{org17}Departamento de F\'{\i}sica de Part\'{\i}culas and IGFAE, Universidad de Santiago de Compostela, Santiago de Compostela, Spain
\item \Idef{org18}Department of Physics and Technology, University of Bergen, Bergen, Norway
\item \Idef{org19}Department of Physics, Aligarh Muslim University, Aligarh, India
\item \Idef{org20}Department of Physics, Ohio State University, Columbus, Ohio, United States
\item \Idef{org21}Department of Physics, Sejong University, Seoul, South Korea
\item \Idef{org22}Department of Physics, University of Oslo, Oslo, Norway
\item \Idef{org23}Dipartimento di Elettrotecnica ed Elettronica del Politecnico, Bari, Italy
\item \Idef{org24}Dipartimento di Fisica dell'Universit\`{a} 'La Sapienza' and Sezione INFN Rome, Italy
\item \Idef{org25}Dipartimento di Fisica dell'Universit\`{a} and Sezione INFN, Cagliari, Italy
\item \Idef{org26}Dipartimento di Fisica dell'Universit\`{a} and Sezione INFN, Trieste, Italy
\item \Idef{org27}Dipartimento di Fisica dell'Universit\`{a} and Sezione INFN, Turin, Italy
\item \Idef{org28}Dipartimento di Fisica e Astronomia dell'Universit\`{a} and Sezione INFN, Bologna, Italy
\item \Idef{org29}Dipartimento di Fisica e Astronomia dell'Universit\`{a} and Sezione INFN, Catania, Italy
\item \Idef{org30}Dipartimento di Fisica e Astronomia dell'Universit\`{a} and Sezione INFN, Padova, Italy
\item \Idef{org31}Dipartimento di Fisica `E.R.~Caianiello' dell'Universit\`{a} and Gruppo Collegato INFN, Salerno, Italy
\item \Idef{org32}Dipartimento di Scienze e Innovazione Tecnologica dell'Universit\`{a} del  Piemonte Orientale and Gruppo Collegato INFN, Alessandria, Italy
\item \Idef{org33}Dipartimento Interateneo di Fisica `M.~Merlin' and Sezione INFN, Bari, Italy
\item \Idef{org34}Division of Experimental High Energy Physics, University of Lund, Lund, Sweden
\item \Idef{org35}Eberhard Karls Universit\"{a}t T\"{u}bingen, T\"{u}bingen, Germany
\item \Idef{org36}European Organization for Nuclear Research (CERN), Geneva, Switzerland
\item \Idef{org37}Excellence Cluster Universe, Technische Universit\"{a}t M\"{u}nchen, Munich, Germany
\item \Idef{org38}Faculty of Engineering, Bergen University College, Bergen, Norway
\item \Idef{org39}Faculty of Mathematics, Physics and Informatics, Comenius University, Bratislava, Slovakia
\item \Idef{org40}Faculty of Nuclear Sciences and Physical Engineering, Czech Technical University in Prague, Prague, Czech Republic
\item \Idef{org41}Faculty of Science, P.J.~\v{S}af\'{a}rik University, Ko\v{s}ice, Slovakia
\item \Idef{org42}Faculty of Technology, Buskerud and Vestfold University College, Vestfold, Norway
\item \Idef{org43}Frankfurt Institute for Advanced Studies, Johann Wolfgang Goethe-Universit\"{a}t Frankfurt, Frankfurt, Germany
\item \Idef{org44}Gangneung-Wonju National University, Gangneung, South Korea
\item \Idef{org45}Gauhati University, Department of Physics, Guwahati, India
\item \Idef{org46}Helsinki Institute of Physics (HIP), Helsinki, Finland
\item \Idef{org47}Hiroshima University, Hiroshima, Japan
\item \Idef{org48}Indian Institute of Technology Bombay (IIT), Mumbai, India
\item \Idef{org49}Indian Institute of Technology Indore, Indore (IITI), India
\item \Idef{org50}Inha University, Incheon, South Korea
\item \Idef{org51}Institut de Physique Nucl\'eaire d'Orsay (IPNO), Universit\'e Paris-Sud, CNRS-IN2P3, Orsay, France
\item \Idef{org52}Institut f\"{u}r Informatik, Johann Wolfgang Goethe-Universit\"{a}t Frankfurt, Frankfurt, Germany
\item \Idef{org53}Institut f\"{u}r Kernphysik, Johann Wolfgang Goethe-Universit\"{a}t Frankfurt, Frankfurt, Germany
\item \Idef{org54}Institut f\"{u}r Kernphysik, Westf\"{a}lische Wilhelms-Universit\"{a}t M\"{u}nster, M\"{u}nster, Germany
\item \Idef{org55}Institut Pluridisciplinaire Hubert Curien (IPHC), Universit\'{e} de Strasbourg, CNRS-IN2P3, Strasbourg, France
\item \Idef{org56}Institute for Nuclear Research, Academy of Sciences, Moscow, Russia
\item \Idef{org57}Institute for Subatomic Physics of Utrecht University, Utrecht, Netherlands
\item \Idef{org58}Institute for Theoretical and Experimental Physics, Moscow, Russia
\item \Idef{org59}Institute of Experimental Physics, Slovak Academy of Sciences, Ko\v{s}ice, Slovakia
\item \Idef{org60}Institute of Physics, Academy of Sciences of the Czech Republic, Prague, Czech Republic
\item \Idef{org61}Institute of Physics, Bhubaneswar, India
\item \Idef{org62}Institute of Space Science (ISS), Bucharest, Romania
\item \Idef{org63}Instituto de Ciencias Nucleares, Universidad Nacional Aut\'{o}noma de M\'{e}xico, Mexico City, Mexico
\item \Idef{org64}Instituto de F\'{\i}sica, Universidad Nacional Aut\'{o}noma de M\'{e}xico, Mexico City, Mexico
\item \Idef{org65}iThemba LABS, National Research Foundation, Somerset West, South Africa
\item \Idef{org66}Joint Institute for Nuclear Research (JINR), Dubna, Russia
\item \Idef{org67}Konkuk University, Seoul, South Korea
\item \Idef{org68}Korea Institute of Science and Technology Information, Daejeon, South Korea
\item \Idef{org69}KTO Karatay University, Konya, Turkey
\item \Idef{org70}Laboratoire de Physique Corpusculaire (LPC), Clermont Universit\'{e}, Universit\'{e} Blaise Pascal, CNRS--IN2P3, Clermont-Ferrand, France
\item \Idef{org71}Laboratoire de Physique Subatomique et de Cosmologie, Universit\'{e} Grenoble-Alpes, CNRS-IN2P3, Grenoble, France
\item \Idef{org72}Laboratori Nazionali di Frascati, INFN, Frascati, Italy
\item \Idef{org73}Laboratori Nazionali di Legnaro, INFN, Legnaro, Italy
\item \Idef{org74}Lawrence Berkeley National Laboratory, Berkeley, California, United States
\item \Idef{org75}Lawrence Livermore National Laboratory, Livermore, California, United States
\item \Idef{org76}Moscow Engineering Physics Institute, Moscow, Russia
\item \Idef{org77}National Centre for Nuclear Studies, Warsaw, Poland
\item \Idef{org78}National Institute for Physics and Nuclear Engineering, Bucharest, Romania
\item \Idef{org79}National Institute of Science Education and Research, Bhubaneswar, India
\item \Idef{org80}Niels Bohr Institute, University of Copenhagen, Copenhagen, Denmark
\item \Idef{org81}Nikhef, National Institute for Subatomic Physics, Amsterdam, Netherlands
\item \Idef{org82}Nuclear Physics Group, STFC Daresbury Laboratory, Daresbury, United Kingdom
\item \Idef{org83}Nuclear Physics Institute, Academy of Sciences of the Czech Republic, \v{R}e\v{z} u Prahy, Czech Republic
\item \Idef{org84}Oak Ridge National Laboratory, Oak Ridge, Tennessee, United States
\item \Idef{org85}Petersburg Nuclear Physics Institute, Gatchina, Russia
\item \Idef{org86}Physics Department, Creighton University, Omaha, Nebraska, United States
\item \Idef{org87}Physics Department, Panjab University, Chandigarh, India
\item \Idef{org88}Physics Department, University of Athens, Athens, Greece
\item \Idef{org89}Physics Department, University of Cape Town, Cape Town, South Africa
\item \Idef{org90}Physics Department, University of Jammu, Jammu, India
\item \Idef{org91}Physics Department, University of Rajasthan, Jaipur, India
\item \Idef{org92}Physik Department, Technische Universit\"{a}t M\"{u}nchen, Munich, Germany
\item \Idef{org93}Physikalisches Institut, Ruprecht-Karls-Universit\"{a}t Heidelberg, Heidelberg, Germany
\item \Idef{org94}Politecnico di Torino, Turin, Italy
\item \Idef{org95}Purdue University, West Lafayette, Indiana, United States
\item \Idef{org96}Pusan National University, Pusan, South Korea
\item \Idef{org97}Research Division and ExtreMe Matter Institute EMMI, GSI Helmholtzzentrum f\"ur Schwerionenforschung, Darmstadt, Germany
\item \Idef{org98}Rudjer Bo\v{s}kovi\'{c} Institute, Zagreb, Croatia
\item \Idef{org99}Russian Federal Nuclear Center (VNIIEF), Sarov, Russia
\item \Idef{org100}Russian Research Centre Kurchatov Institute, Moscow, Russia
\item \Idef{org101}Saha Institute of Nuclear Physics, Kolkata, India
\item \Idef{org102}School of Physics and Astronomy, University of Birmingham, Birmingham, United Kingdom
\item \Idef{org103}Secci\'{o}n F\'{\i}sica, Departamento de Ciencias, Pontificia Universidad Cat\'{o}lica del Per\'{u}, Lima, Peru
\item \Idef{org104}Sezione INFN, Bari, Italy
\item \Idef{org105}Sezione INFN, Bologna, Italy
\item \Idef{org106}Sezione INFN, Cagliari, Italy
\item \Idef{org107}Sezione INFN, Catania, Italy
\item \Idef{org108}Sezione INFN, Padova, Italy
\item \Idef{org109}Sezione INFN, Rome, Italy
\item \Idef{org110}Sezione INFN, Trieste, Italy
\item \Idef{org111}Sezione INFN, Turin, Italy
\item \Idef{org112}SSC IHEP of NRC Kurchatov institute, Protvino, Russia
\item \Idef{org113}SUBATECH, Ecole des Mines de Nantes, Universit\'{e} de Nantes, CNRS-IN2P3, Nantes, France
\item \Idef{org114}Suranaree University of Technology, Nakhon Ratchasima, Thailand
\item \Idef{org115}Technical University of Split FESB, Split, Croatia
\item \Idef{org116}The Henryk Niewodniczanski Institute of Nuclear Physics, Polish Academy of Sciences, Cracow, Poland
\item \Idef{org117}The University of Texas at Austin, Physics Department, Austin, Texas, USA
\item \Idef{org118}Universidad Aut\'{o}noma de Sinaloa, Culiac\'{a}n, Mexico
\item \Idef{org119}Universidade de S\~{a}o Paulo (USP), S\~{a}o Paulo, Brazil
\item \Idef{org120}Universidade Estadual de Campinas (UNICAMP), Campinas, Brazil
\item \Idef{org121}University of Houston, Houston, Texas, United States
\item \Idef{org122}University of Jyv\"{a}skyl\"{a}, Jyv\"{a}skyl\"{a}, Finland
\item \Idef{org123}University of Liverpool, Liverpool, United Kingdom
\item \Idef{org124}University of Tennessee, Knoxville, Tennessee, United States
\item \Idef{org125}University of the Witwatersrand, Johannesburg, South Africa
\item \Idef{org126}University of Tokyo, Tokyo, Japan
\item \Idef{org127}University of Tsukuba, Tsukuba, Japan
\item \Idef{org128}University of Zagreb, Zagreb, Croatia
\item \Idef{org129}Universit\'{e} de Lyon, Universit\'{e} Lyon 1, CNRS/IN2P3, IPN-Lyon, Villeurbanne, France
\item \Idef{org130}V.~Fock Institute for Physics, St. Petersburg State University, St. Petersburg, Russia
\item \Idef{org131}Variable Energy Cyclotron Centre, Kolkata, India
\item \Idef{org132}Vin\v{c}a Institute of Nuclear Sciences, Belgrade, Serbia
\item \Idef{org133}Warsaw University of Technology, Warsaw, Poland
\item \Idef{org134}Wayne State University, Detroit, Michigan, United States
\item \Idef{org135}Wigner Research Centre for Physics, Hungarian Academy of Sciences, Budapest, Hungary
\item \Idef{org136}Yale University, New Haven, Connecticut, United States
\item \Idef{org137}Yonsei University, Seoul, South Korea
\item \Idef{org138}Zentrum f\"{u}r Technologietransfer und Telekommunikation (ZTT), Fachhochschule Worms, Worms, Germany
\end{Authlist}
\endgroup

\end{document}